\documentclass[10pt,journal]{IEEEtran}

\usepackage{tabu}
\usepackage{outline}
\usepackage{pmgraph}
\usepackage[normalem]{ulem}
\usepackage[utf8]{inputenc}
\usepackage{amssymb}
\usepackage{hyperref}
\usepackage{amsmath}
\usepackage{graphicx}
\usepackage{times}
\usepackage{xcolor}
\usepackage{xspace}
\usepackage[colorinlistoftodos]{todonotes} 
\usepackage{cite}
\usepackage{bm} 
\usepackage{url}

\usepackage{epstopdf}
\AppendGraphicsExtensions{.tiff}
\graphicspath{{fig/}} 

\usepackage{epsfig}
\usepackage{tikz}
\usetikzlibrary{spy}
\usepackage{algpseudocode}
\usepackage{algorithm}
\usepackage{mathrsfs}

\def\QED{~\rule[-1pt]{5pt}{5pt}\par\medskip}

\long\def\comment#1{} 

\newcommand{\xmath}[1] {\ensuremath{#1}\xspace}
\newcommand{\blmath}[1] {\xmath{\bm{#1}}}

\newcommand{\B}{\blmath{B}}

\newcommand{\E}{\blmath{E}}

\newcommand{\z}{\blmath{z}}

\newcommand{\Bb}{{\blmath B}}

\newcommand{\Db}{{\blmath D}}
\newcommand{\Eb}{{\blmath E}}

\newcommand{\Ib}{{\blmath I}}

\newcommand{\Rb}{{\blmath R}}

\newcommand{\Wb}{{\blmath W}}
\newcommand{\Xb}{{\blmath X}}
\newcommand{\Yb}{{\blmath Y}}
\newcommand{\Zb}{{\blmath Z}}

\renewcommand{\sb}{{\blmath s}}

\newcommand{\ub}{{\blmath u}}
\newcommand{\vb}{{\blmath v}}
\newcommand{\wb}{{\blmath w}}
\newcommand{\xb}{{\blmath x}}
\newcommand{\yb}{{\blmath y}}
\newcommand{\zb}{{\blmath z}}

\newcommand{\Tc}{\mathcal{T}}

\newcommand{\Phib}{{\boldsymbol {\Phi}}}

\newcommand{\Thetab}{{\boldsymbol {\Theta}}}

\newcommand{\Rd}{{\mathbb R}}

\newcommand{\psib}{{\boldsymbol{\psi}}}

\newcommand{\chib}{{\boldsymbol {\chi}}}

\newcommand{\Ec}{{{\mathcal E}}}
\newcommand{\Dc}{{{\mathcal D}}}

\newcommand{\beq}{\begin{equation}}
\newcommand{\eeq}{\end{equation}}
\newcommand{\beqa}{\begin{eqnarray}}
\newcommand{\eeqa}{\end{eqnarray}}

\newcommand{\Lambdab}{\boldsymbol{\Lambda}}

\begin{document}

\title{Adaptive and Compressive Beamforming Using Deep Learning for Medical Ultrasound}
\author{
Shujaat Khan,
        Jaeyoung Huh,~
        and~Jong~Chul~Ye,~\IEEEmembership{Fellow,~IEEE}
\thanks{This work was supported by the National Research Foundation (NRF) of Korea grant NRF-2016R1A2B3008104.The authors are with the Department of Bio and Brain Engineering, Korea Advanced Institute of Science and Technology (KAIST), Daejeon 34141, Republic of Korea (e-mail:\{shujaat,woori93,jong.ye\}@kaist.ac.kr). 
		}}

\maketitle

\begin{abstract}
In ultrasound (US) imaging, various types of adaptive beamforming techniques have been investigated to improve the resolution and  contrast-to-noise ratio of the delay and sum (DAS) beamformers. Unfortunately, the performance of these adaptive beamforming approaches degrade when the underlying model is not sufficiently accurate and the number of channels decreases. To address this problem, here we propose a deep learning-based beamformer to generate significantly improved  images  over widely varying measurement conditions and  channel subsampling patterns. In particular, our deep neural network is designed to directly process full or sub-sampled radio-frequency (RF) data acquired at various subsampling  rates and detector configurations so that it can generate high quality ultrasound images  using a single beamformer. The origin of such input-dependent adaptivity is also theoretically analyzed. Experimental results using B-mode focused ultrasound  confirm the efficacy of the proposed methods.
\end{abstract}

\begin{IEEEkeywords}
Ultrasound imaging, B-mode, beamforming, adaptive beamformer, Capon beamformer
\end{IEEEkeywords}

\IEEEpeerreviewmaketitle

\section{Introduction}
\label{sec:introduction}

Excellent temporal resolution with reasonable image quality makes the ultrasound (US) modality a first choice for variety of clinical applications. Moreover, due to its non-invasive nature, 
US  is an indispensable tool for  some clinical applications such as cardiac,  fetal imaging, etc. 

In US, an image reconstruction is usually done by back-propagating the preprocessed measurement data  and adding all the contributions.
For example, in focused B-mode US imaging, the return echoes from individual scan lines are recorded by the receiver channels (Rx),
after which  a delay and sum (DAS) beamformer applies appropriate time-delays  to the channel measurements and additively
combines them for each depth
to form an image at each scan line.
Despite the simplicity,   large number of receiver elements are
often necessary in DAS beamformer to improve the image quality by reducing the side lobes.
Moreover, to calculate accurate time delay, sufficiently large bandwidth transducers are required.

To deal with unfavorable acquisition conditions,
 various adaptive beamforming techniques have been developed over the several decades \cite{AdaptiveBF1,CaponBF,CaponBF2,MVBF1,MVBF2,BSBF1,FastRobustBF,MultiBeamBF1,IterativeBF1}.
The main idea of adaptive beamforming is to change the receive aperture weights based on the received data statistics to improve the resolution
and  enhance the contrast.
 One of the most extensively studied adaptive beamforming techniques is Capon beamforming, also known
 as the minimum variance (MV) beamforming \cite{CaponBF,CaponBF2,MVBF1}. The aperture weight of Capon beamfomer is
 derived by minimizing the side lobe while maintaining the gain in the look-ahead direction.
 Unfortunately,  Capon beamforming is  computationally heavy for practical use due to
  the calculation of the spatial covariance matrix of channel data and its inverse \cite{MVBF2}. Moreover,
  the performance of Capon beamformer is dependent upon the accuracy of the covariance matrix estimate.
To address these problems, many improved versions of MV beamformers have been proposed \cite{MVBF1,MVBF2,BSBF1,FastRobustBF}.
Some of the notable examples include the beamspace adaptive beamformer \cite{BSBF1},
multi-beam Capon based on multibeam covariance matrices\cite{MultiBeamBF1}, etc.
In addition, a parametric form of iterative update covariance matrix calculation has been proposed instead of calculating the empirical covariance matrix \cite{IterativeBF1}.

On the other hand, compressive beamforming method have been also extensively investigated  to reduce the data rate \cite{jin2016compressive}.
Specifically, high-quality ultrasound imaging demands for significantly high sampling rates, which eventually increases the volume of data transmitted from the system’s front end. Moreover, in 3-D ultrasound imaging, 2-D transducer arrays are used and more scan lines are needed, which leads to vastly increased amount of sampled data  with respect to 2-D imaging.
To achieve the aforementioned data rate reduction, random RF sub-sampling  has been employed in various ultrasound imaging  researches, e.g. \cite{schretter2017ultrasound}, etc. 
Many researches also suggested the buffered probe sampling which can reduce the number of scan lines at the cost of complexity in probe design. 
Then specially designed compressive beamforming techniques were used to exploit the redundancy in the image
to compensate for the reduced measurement data \cite{7420721,jin2016compressive}. Unfortunately, most of the 
existing compressive beamformers require either hardware changes \cite{wagner2011xampling} or computationally
expensive optimization methods \cite{jin2016compressive}.

Recently, inspired by the tremendous success of deep learning,
many researchers have investigated deep learning approaches for various inverse problems \cite{kang2017deep,kang2018deep,chen2017lowBOE,adler2018learned,wolterink2017generative,jin2017deep,han2017framing,wang2016accelerating,hammernik2018learning,schlemper2018deep,zhu2018image,lee2018deep}.
In US literature, the works in \cite{Allman_reviewer,luchies2018deep} were among the first to apply deep learning approaches to US image reconstruction.  In particular, Allman \textit{et al} \cite{Allman_reviewer} proposed a machine learning method to identify and remove reflection artifacts in photo-acoustic channel data.  Luchies and Byram \cite{luchies2018deep} proposed a frequency domain deep learning method
for suppressing off-axis scattering in
ultrasound channel data.
 In \cite{feigin2018deep}, a deep neural network is designed to estimate the attenuation characteristics of sound in human body. In \cite{perdios2017deep,zhou2018high}, ultrasound image denoising method is proposed for the B-mode and single angle
 plane wave imaging.
Rather than using  deep neural network as a post processing method, the authors in
 \cite{yoon2018efficient,gasse2017high,MICCAI1,MICCAI2} employed
 deep neural networks for the reconstruction of high-quality US images from limited number of received RF data.   
For example, the work in \cite{gasse2017high} uses deep neural network for coherent compound imaging
from small number of plane wave illumination. 
In focused B-mode ultrasound imaging, \cite{yoon2018efficient} employs the deep neural network
to interpolate the missing RF-channel data with multiline aquisition for accelerated scanning.

While these recent deep neural network approaches provide impressive reconstruction
performance,  the designed
neural network cannot completely replace a DAS beamformer, since they are designed as  pre- or post- processing
steps  for specific acquisition scenarios and many of the works employ the standard DAS beamformer.
Therefore,  one of the most important contributions of this paper is to   replace
the DAS, adaptive, or compressive beamformers with a deep learning-based data-driven adaptive deep beamformer (DeepBF) so that
 a { single}  DeepBF can generate high quality images robustly for
 various detector channel configurations.  Moreover,  unlike the MV beamformer that can be used only for uniform  array,  our DeepBF
is designed for various detectors  and RF subsampling schemes,  in spite of significantly
reduced run-time computational complexity.
In contrast to  \cite{yoon2018efficient}, where the deep learning approach was developed to interpolate missing RF data to be used as input to the standard beamformer,  the proposed method is a CNN-based beamforming pipeline, without requiring additional beamformer.  Consequently, this approach is much simpler and can be easily incorporated to replace the standard beamforming pipeline.  
 Despite the simplicity,   our experiments show that direct reconstruction using 
 the proposed DeepBF  produces better results compared to \cite{yoon2018efficient}.

The consistent performance improvement over widely varying subsampling rates using a single CNN may appear mysterious.
Inspired by the recent theoretical understanding  of deep convolutional framelets \cite{ye2017deep,ye2019understanding},
another important contribution of  this paper is a detailed theoretical analysis to identify the origin
of  the input adaptivity and the performance improvement of DeepBF. Our theoretical analysis suggests that the deep learning-based beamformer
may be the right direction for medical ultrasound.

 After the initial work of this paper became available on Arxiv \cite{khan2019universal},  a related work on deep learning based adaptive
 beamformer appeared \cite{8683478}. In contrast to the proposed method, \cite{8683478} is interested in estimating the adaptive beamformer
 weights using a deep neural network. Moreover,  the results are only available for simple phantom data, the application of compressive
 beamforming was not considered,  and the theoretical
 analysis to unveil why the deep learning beamformer works was not provided. Therefore, our work is more general and provides a systematic understanding in
 designing deep learning based beamformer.

This paper is  organized as follows. In Section~\ref{sec:review},  a brief survey of the existing adaptive beamforming methods are provided, which is followed by
the detailed explanation of the proposed deep beamformer in Section~\ref{sec:theory}.
Section~\ref{sec:methods} then describes the data set and experimental setup.
Experimental results are provided in Section~\ref{sec:results}, which is followed
by Discussion and Conclusions in Section~\ref{sec:discussion} and Section~\ref{sec:conclusion}, respectively.

\begin{figure*}[!t]
   \centering
   \centerline{\epsfig{figure=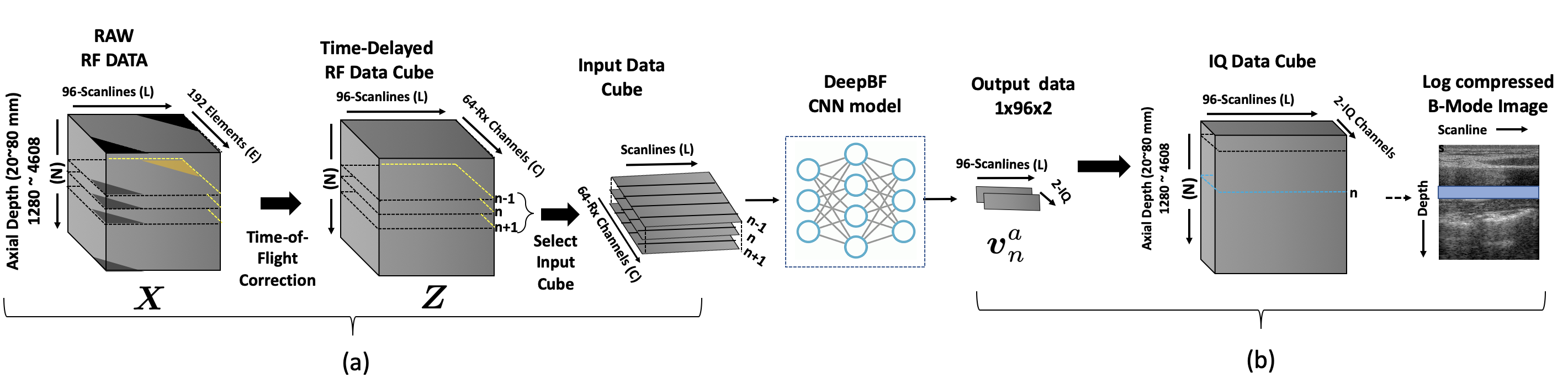, width=18cm}}   
    \caption{Our deep beamformer pipeline.  The highlighted region in B-mode image is the  output generated by the neural network
    for given input RF data.}
    \label{fig:system_block_diagram}
\end{figure*}

\section{Mathematical Preliminaries}
\label{sec:review}


\subsection{Notation}

In this paper, the uppercase boldface  letter such as $\Xb,\Yb_n$ are used to refer  matrices {and} tensors, whereas
the lowercase boldface letters such as $\xb,\yb_{l,n}$ represent the {vectors}.  Non-bold letters such as $x,v_{l,n}$ denote scalars.

Measured RF data  is a three-dimensional cube $\Xb \in \Rd^{L\times N\times E}$ from the B-mode ultrasound as shown in Fig.~\ref{fig:system_block_diagram} (a),
where $L, N,$ and $E$ denote the number of scan lines (or transmit events (TE)), depth planes, and the number of probe elements, respectively.
The RF data cube $\Xb$ is often represented as $\Xb:=[\xb_{l,n}]_{l,n}$, where
$\xb_{l,n}\in \Rd^{E}$ is the $(l,n)$-th element of the data cube, representing  the RF data measured by the receiver channels  from the the $l$-th scan line at the depth index $n$.
The time-delay corrected data cube $\Yb \in \Rd^{L\times N\times E}$ is similarly denoted by
$\Yb=[\yb_{l,n}]_{l,n}$, where
 \begin{eqnarray}
 \yb_{l,n}=\begin{bmatrix} y_{l,n}[0] & y_{l,n}[1] & \cdots &y_{l,n}[E-1]\end{bmatrix}^\top \in \Rd^{E}
 \end{eqnarray}
with
\begin{align*}
{	y_{l,n}[i]:=x_{l,n-\tau_{l,n}[i]}[i]}
\end{align*}
where {$^\top$ denotes the transpose,} $\tau_{l,n}[i]$ is the time delay for the
 $i$-th receiver elements to obtain the $l$-th  scan line at the depth  $n$.

 In many US imaging,  only  a subset of receiver channels are used to process return echoes to save
 power consumption and/or data rate. Usually, the aperture, which refers to the span of the active receiver,
 varies depending on the scan lines, so that symmetric set of receivers along the scan lines are used.
 In this case, the received RF data can be explicitly modeled as
 $\Zb=[\zb_{l,n}]_{l,n}$, where
 \begin{eqnarray}
 \zb_{l,n}=\begin{bmatrix} z_{l,n}[0] & z_{l,n}[1] & \cdots &z_{l,n}[{C}-1]\end{bmatrix}^\top \in \Rd^{C}
 \end{eqnarray}
and
 \begin{align}\label{eq:z}
 z_{l,n}[i]= y_{l,n}[i+d_l]
 \end{align}
 where $d_l$ denotes the specific detector offset to indicate the active channel elements, which is determined
 for each scan line index $l$, {and $C$ is the aperture size.}  See Fig.~\ref{fig:system_block_diagram}(a) for the conversion between the data cube $\Xb$ and $\Zb$,
 where the dark triangular  regions in $\Xb$, which correspond to inactive receiver elements, are removed in constructing $\Zb$.
%

\subsection{Classical Beamforming}

\subsubsection{DAS beamforming}

The standard  delay and sum (DAS) beamformer for the $l$-th scanline at the depth sample $n$ can be expressed as
\begin{equation}\label{eq:DAS}
{u}_{l,n} 
=\frac{1}{J}\mathbf{1}^\top\zb_{l,n} 
\end{equation}
where $\mathbf{1}$ denotes a $C$-dimensional column-vector of ones, and $J$ is the number of active {channels}.

 \subsubsection{Adaptive beamforming}

The DAS beamformer is designed to extract the  low-frequency spatial content that corresponds to the energy within the main lobe; thus,
it is difficult to control side lobe leakage. Reduced side lobe leakage can be achieved by replacing the uniform weights by tapered weights:
\begin{equation}\label{BF2}
u_{l,n}  = \wb_{l,n}^\top \zb_{l,n},
\end{equation}
where
$\wb_{l,n} \in \Rd^C.$
Specifically, in adaptive beamforming the objective is to find the $\wb_{l,n}$ that minimizes the variance of ${u}_{l,n}$, subject to the constraint that the gain in the desired beam direction equals unity. For example, the minimum variance (MV) estimation task can be formulated as  \cite{CaponBF,CaponBF2,MVBF1}

\begin{equation*}
\begin{aligned}
& \underset{\wb_{l,n}}{\text{minimize}}
& & E[|u_{l,n}]|^2]=\min_{\wb_{l,n}}\wb_{l,n}^{\top}\Rb_{l,n}\wb_{l,n} \\
& \text{subject to}
& &   \mathbf{1}^\top \wb_{l,n} =1,
\end{aligned}
\end{equation*}
where $E[\cdot]$ is the expectation operator over RF data distribution, and $\Rb_{l,n}$ is a spatial covariance matrix given by:
\begin{equation}
\Rb_{l,n}=E\left[\zb_{l,n} \zb_{l,n}^{\top}\right] .
\end{equation}
Then, $\wb_{l,n}$ can be obtained by  Lagrange multiplier method \cite{LagrangeBF} and expressed as
\begin{equation}\label{eq:wb}
\wb_{l,n}=\frac{\Rb_{l,n}^{-1} \mathbf{1}}{\mathbf{1}^{\top} \Rb_{l,n}^{-1} \mathbf{1}} \quad  .
\end{equation}

\subsection{Deconvolution Ultrasound}

One of the main limitations of the aforementioned beamforming methods is that they 
 are based on the ray approximation of the
wave propagation, whereas
 the real sound propagations exhibits many wave phenomenon such as scattering, {diffraction,} etc.
Moreover, the precision of  the time delay  $\tau_{l,n}[i]$ calculation is determined by bandwidth of the transducers, which limits
 the accuracy of delayed signal $y_{l,n}[i]:=x_{l,n}\left[i-\tau_{l,n}[i]\right] $.
These modeling {inaccuracies} may affect
the spatial resolution and the contrast of standard US images. 

In order to overcome these issues, many researchers have
explored the deconvolution of US images \cite{chen2015compressive,jensen1992deconvolution}.
Specifically,  the deconvolution US tries to find the  filter kernel $h_{l,n}$ such that the filtered
signal $v_{l,n}$ given by
\begin{eqnarray}
v_{l,n} &=& (h\ast u)_{l,n}
\label{eq:deconv}
\end{eqnarray}
produces high resolution images.

\subsection{Imposing Causality Condition}
Another important step after the beamforming is to convert the processed data to a causal signal  using
 Kramers–Kronig relation \cite{o1981kramers}.  This step is necessary to detect the signal envelope.
More specifically,  this process is performed by
\begin{eqnarray}\label{eq:kramer}
v_{l,n}^a= v_{l,n}+ \iota (\kappa \ast v)_{l,n} = \left((\delta +\iota\kappa) \ast v\right)_{l,n}  ,%
\end{eqnarray}
where $\iota = \sqrt{-1}$, and $\delta$ is a discrete Dirac delta function,
and  $\kappa$ denotes the filter kernel for Hilbert transform. The filter kernel for Hilbert transform is in principle one-dimensional
since it is applied along the
depth direction.  Here, $v_l^a[n]$ is often referred to as the in-phase (I) and quadrature (Q) representation.

\subsection{Putting Together}

By  using  \eqref{BF2}, \eqref{eq:deconv}  and \eqref{eq:kramer}, we can obtain the following representation: 
\begin{eqnarray}
v_{l,n}^a &=& \left((h+\iota \kappa \ast h )\ast u\right)_{l,n} 
\end{eqnarray} 
If we  define
\begin{align}\label{eq:vba}
\vb^a = \begin{bmatrix} \vb_0^a \\ \vdots \\ \vb_{N-1}^a \end{bmatrix},&\quad \mbox{where}\quad \vb_n^a = \begin{bmatrix} v_{0,n}^a \\ \vdots \\ v_{L-1,n}^a  \end{bmatrix}
\end{align}
and
\begin{align}\label{eq:zb}
\zb = \begin{bmatrix} \zb_0 \\ \vdots \\ \zb_{N-1} \end{bmatrix},&\quad \mbox{where}\quad \zb_n = \begin{bmatrix} z_{0,n} \\ \vdots \\ z_{L-1,n}  \end{bmatrix}
\end{align}
then the following matrix representation can be obtained:
\begin{align}
\vb^a = \tilde\Bb \ub \label{eq:v}
\end{align}
where $\tilde\Bb$
is a 2-D convolution matrix composed of the filter kernel $h+\iota \kappa \ast h$, and
\begin{align*}
\ub :=\Bb(\zb)\zb,& \quad\mbox{where}\quad \Bb(\zb) = \Ib \otimes \Wb(\zb)^\top 
\end{align*}
where $\otimes$ is a Kronecker product and  $\Wb(\zb)$ is the input-dependent
 beamformer adaptive weight matrix given by
\begin{align*}
\Wb(\zb)&:= \begin{bmatrix} \wb_{0,0}  & \cdots & \wb_{L-1,0} & \cdots & \wb_{L-1,N-1} \end{bmatrix}
\end{align*}
Accordingly, Eq.~\eqref{eq:v} can be equivalently represented as a nonlinear mapping:
\begin{eqnarray}\label{eq:BB}
\vb^a &=& \Tc(\zb)\zb 
\end{eqnarray}
where  $\Tc(\zb):=\widetilde\Bb \Bb(\zb).$
Then, the goal of the US reconstruction is to find the nonlinear mapping $\Tc(\zb)$
  so that the processed image has a high resolution with  good contrast  and better signal-to-noise ratio.

\section{Main Contribution}
\label{sec:theory}

\subsection{Piecewise linear approximation  using CNN}

In practice,  the estimation of $\Tc(\zb)$ in \eqref{eq:BB}  is technically challenging.
This is because the beamformer weights  
 are dependent on each RF data $\zb$.
Moreover, the deconvolution filter matrix could be also  spatially varying.
Therefore, the exact calculation is usually computationally expensive. 
A quick remedy to overcome this
would be precalculating nonlinear mapping $\Tc(\zb)$.  Unfortunately,   it requires
huge memory to store $\Tc(\zb)$ for all  $\zb$. 

In this regard,  a convolutional neural network (CNN)  using ReLU nonlinearities provides an ingenious way of  addressing this issue.
Specifically, in our recent theoretical work \cite{ye2019understanding},  we have shown that an encoder-decoder CNN with ReLU nonlinearity 
generates large number of distinct  linear mappings depending on inputs. 
More specifically, the input space 
is partitioned into non-overlapping regions where input for each region share a common linear representation or mapping.
\cite{ye2019understanding}.

\begin{figure}[!hbt]
   \centerline{\epsfig{figure=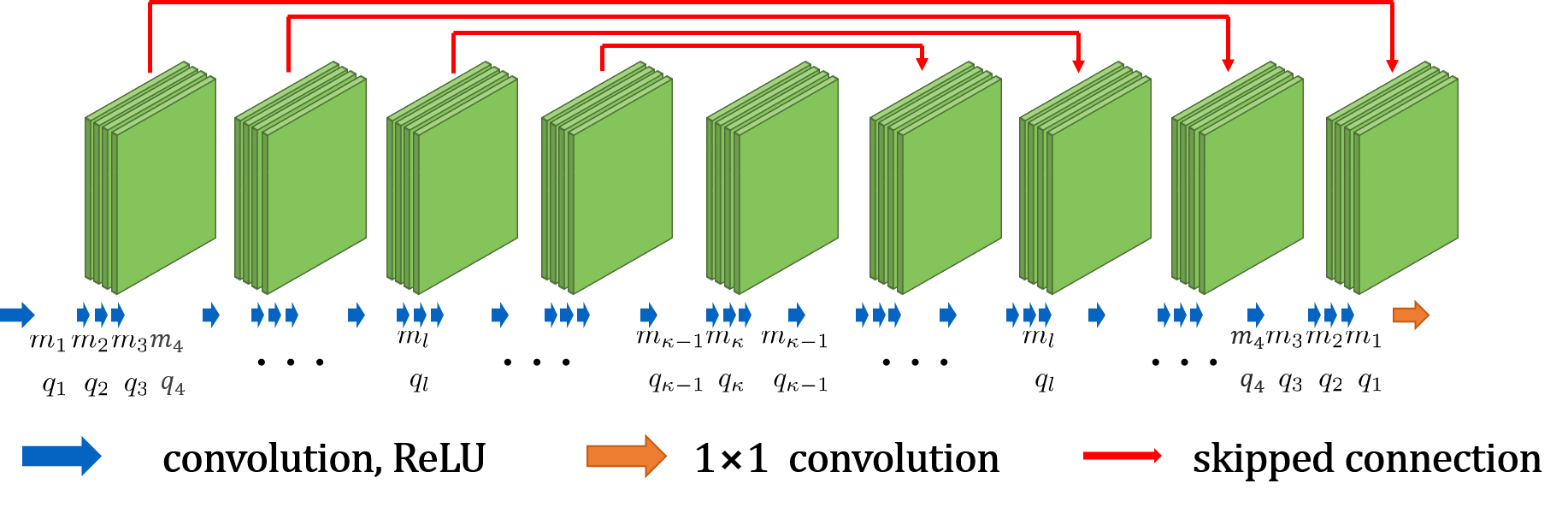, width=9cm}}
   \caption{Encoder-decoder CNN backbone.}
   \label{fig:CNN_block_diagram}
\end{figure}

Specifically, consider an encoder-decoder CNN with the output $\vb^a$  with respect to input $\z$ as shown in Fig.~\ref{fig:CNN_block_diagram}, where
there exists  skipped connection  for every four convolution operations.
As shown in Appendix, 
the output $\vb^a$ of the encoder-decoder CNN  with respect to input $\z$ can be represented by the following nonlinear mapping 
\begin{eqnarray}\label{eq:basis0}
\vb^a 
~= \Tc_\Thetab(\zb)\zb = \sum_{i} \left\langle {\blmath b}_i(\z), \z \right\rangle \tilde  {\blmath b}_i(\z)
\end{eqnarray}
where  $\Thetab$ refers to all the convolution filter parameters,
and $ {\blmath b}_i(\z)$ and $\tilde  {\blmath b}_i(\z)$ denote the $i$-th column of the matrices \eqref{eq:Bz} and \eqref{eq:tBz}, respectively.

The expression \eqref{eq:basis} 
reveals many important aspects of neural networks.
First,  
the CNN representation in \eqref{eq:basis} has explicit dependency   on the input $\z$  in \eqref{eq:basis}, due to the input
dependent ReLU activation pattern.
Accordingly, even from the same filter set,
the input-dependent ReLU activation pattern makes the resulting mapping vary depending
on the input signals.  Furthermore,  the number of distinct
linear representation increases exponentially with the number of neurons  determined by network depth  and width \cite{ye2019understanding}, since  the distinct ReLU activation pattern is  in principle combinatorially many  up to $2^{\text{\# of neurons}}$.
Second, the number of blocks in $\B(\z)$ and $\tilde \B(\z)$ in \eqref{eq:Bz} and \eqref{eq:tBz} are determined by the number of skipped connections,
so  the skipped branch makes the representation more redundant, which again makes the
neural network have more piecewise linear regions \cite{ye2019understanding}.

Note that the piecewise linear representation using \eqref{eq:basis0}  is useful for approximating
our nonlinear mapping $\Tc(\zb)$  in \eqref{eq:BB}.
Specifically, the piecewise linear representation by the DeepBF can be obtained by learning  the filters $\Thetab$ from the following
 optimization problem:
\begin{eqnarray}\label{eq:train_org}
\min_{\Thetab} \sum_{t=1}^T\|\vb^{a(t)} - \Tc_\Thetab \zb^{(t)} \|_2^2,
\end{eqnarray}
where $\{(\zb^{(t)},\vb^{a(t)})\}_{t=1}^T$ denotes the training data set composed of  RF data and the target IQ data, which are collected across
all subjects and subsampling patterns. Although the piecewise linear property of CNN  may appear as a limitation to approximate arbitrary nonlinear functions,
it also provides good architectural prior known as {\em inductive bias,} which results in inherent regularization effects.
As will be shown in experimental results, we found that this inductive bias works {favorably} for US reconstruction.

\subsection{Proposed Deep Beamformer Pipeline}

One of the limitations of the original training framework in \eqref{eq:train_org} is that the RF data input 
$\zb$ and the target $\vb^a$ requires too big memory to store using GPU memory. Therefore,
rather than training a neural network to learn the mapping between all RF data,  we implemented
a separable form of a neural network such that the neural network is trained to estimate one depth at a time.
Specifically, with a slight abuse of notation, our neural network is designed as
\begin{eqnarray}\label{eq:our}
\vb_n^a 
~= \Tc_\Thetab\left(\sb_n\right)\sb_n,&&
\end{eqnarray}
where $\sb_n$ is a sub-set of input RF data $\zb$ collected from three depth planes around the depth $n$:
 $$\sb_n = \begin{bmatrix}\zb_{n-1}^\top & \zb_n^\top & \zb_{n+1}^\top \end{bmatrix}^\top$$
 where $\vb_n^a$ and $\zb_n$ are defined in \eqref{eq:vba} and \eqref{eq:zb}, respectively.
Then, the resulting neural network training is given by
 \begin{eqnarray}\label{eq:train_new}
\min_{\Thetab} \sum_{t=1}^T\sum_{n=1}^N\|\vb_n^{a(t)} - \Tc_\Thetab \sb_n^{(t)} \|_2^2,
\end{eqnarray}
where $\{(\sb_n^{(t)},\vb_n^{a(t)})\}_{n,t=1}^{N,T}$ denotes the training data set  collected across all depth planes.

One of the potential limitations of using the  restricted architecture \eqref{eq:our}  is  the reduction of the depth-dependent adaptivity of the
neural network, since all the depth information 
is used together as a target data. This issue will be revisited in experimental section.

 Fig.~\ref{fig:system_block_diagram} illustrates the proposed DeepBF pipleline using
 the reflected sound waves in the medium measured by the transducer elements. 
{ As a preprocessing for DeepBF pipeline, each measured RF signal $\Xb$ is time-delayed to generate
focused RF data cube $\Zb$
 based on the traveled distance. 
 Then, our DeepBF generates IQ data $\vb^a$ directly from the  time delayed RF data.}
 Compared to the standard DAS beamformer,  this corresponds to the replacement of the deconvolution,
 beamforming and Hilbert transform parts with a deep neural network.
%
Then,  the signal envelope is generated by calculating the sum of squares of 
 the in-phase and quadrature phase signals generated from  the Hilbert transform. 
 Finally, log compression is applied to generate the B-mode images.

\section{Method}
\label{sec:methods}

\subsection{Dataset}

For experimental verification, we used an E-CUBE 12R US system (Alpinion Co., Korea).  For data acquisition, we used a linear array transducer (L3-12H), whose configuration is given in Table \ref{probe_config}.
Specifically, using the linear probe with a center frequency of $8.5$ MHz, we acquired RF data from the carotid/thyroid area of $10$ volunteers {using focused B-mode US imaging}.  The \textit{in vivo} data consist of $40$ temporal frames per subject, providing $400$ sets of TE-Depth-Rx data cube {$\Xb$}. In addition, we acquired $218$ frames of RF data from the ATS-539 multipurpose tissue mimicking phantom using $8.5$ MHz center frequency. {The phantom dataset was only used for test purposes and no additional training of CNN was performed on it.}
{In addition to the carotid/thyroid and phantom datasets we also acquired {datasets} from forearm and calf muscles. In particular, they were acquired using $10$ MHz carrier frequency and consist of $100$ frames $50$ from each body part. {These} data set were used to further validate the generalization power of our trained model,} {and no additional training of CNN was performed on it.}

{For all scans the axial depth was in the range of 20$\sim$80 mm, while lateral length was 38.4mm. Depending on the object of interest, the focal depth is {adjusted} accordingly, in particular it varies in the range of 10$\sim$40 mm. 
}

\begin{table}[!hbt]
	\centering
	\caption{Probe Configuration}
	\label{probe_config}
	\resizebox{0.35\textwidth}{!}{
	\begin{tabular}{c|c}
\hline
		{Parameter} & {Linear Probe} \\ \hline\hline
		Probe Model No.& L3-12H \\
		Carrier wave frequency & 8.5/10.0 MHz\\
		Sampling frequency & 40 MHz \\
		{Scan wave mode} & {Focused} \\
		No. of probe elements & 192 \\
		No. of Tx elements & 128 \\
		No. of TE events & 96\\
		No. of Rx elements & 64 {(from center of Tx)}\\
		Elements pitch & 0.2 mm\\
		Elements width & 0.14 mm \\
		Elevating length & 4.5 mm\\  \hline
	\end{tabular}}
\end{table}

\subsection{Network specification}

Fig.~\ref{fig:CNN_block_diagram} illustrates the schematic diagram of our deep beamformer.
One minor improvement is the channel augmentation at the skipped branch of the decoder rather than simple addition.
{Moreover, to make the neural network process only real-valued data,
the real and image components of $\vb^a$ are separately processed to generate two  channel IQ output.}

{The proposed CNN consists of $37$ convolution layers  (i.e. $\kappa=4$) composed of a contracting path with concatenation, batch normalization, and ReLUs except for the last convolution layer. The first $36$ convolution layers as shown in Fig.~\ref{fig:CNN_block_diagram} use $3\times3$ convolutional filters (i.e., the 2-D filter has a dimension of $3\times 3$), and the last convolution layer uses a $3\times1$ filter followed by an average pooling to contract the $3\times 96\times 64$ data-cube from Depth-TE-Rx sub-space to $1\times96\times2$ Depth-TE-IQ plane. The number of CNN filter channel for each layer is $q_l=64$ and the dimension of the signal is
$m_l=64\times 96$ up to the last layer which shrinks it to $1\times96\times2$ Depth-TE-IQ data.}

\subsection{RF data sampling scheme}
\label{sec:RFsamScon}

{The input and output data configurations are shown in {Fig}.~\ref{fig:system_block_diagram} (a) and (b) respectively. 
The time-delayed RF data cube {$\Zb$} is a three-dimensional data cube composed of total $1280\sim 4608$ depths of
$96\times64$ data in  TE-Rx direction.}
We trained our neural network using multiple input/output pairs, where an input consists of $3\times 96\times 64$ data-cube in the Depth-TE-Rx volume
and  the output is composed of  $96$ pairs of I/Q data in the Depth-TE plane. 
Each target IQ  pair corresponds to two output channels, each representing real and imaginary parts.
{Using carotid/thyroid} dataset, a set of $30,000$ Depth-TE-Rx  cubes of size  $3\times 96\times 64$  were randomly selected from {$10$ frames of} four different subject's datasets, which are divided into $25,000$ {samples} for training and $5,000$ {samples} for validation.  The remaining dataset of $360$ {carotid/thyroid, $218$ phantom  were used as a test data}.
{In addition, to see the generalization capability of the algorithm,
$100$ frame data from totally different anatomical regions (forearm/calf muscles) were} used as a test dataset.

For {compressive beamforming} experiments, in addition to the full RF data with  $64$ RF-channels,
we generated five sets of sub-sampled RF data at different down-sampling rates. More specifically,
the subsampling cases included $32$, $24$, $16$, $8$ and $4$ Rx-channels at {two subsampling schemes, such as 
variable down-sampling patterns across the depth  and  fixed down-sampling patterns across the depth. }
{Since the active receivers at the center of the scan line get RF data from direct reflection, the two channels in the center were always included, the remaining channels are randomly selected from the other $62$ active receiving channels and unselected Rx channels are zero-padded.}
{In variable sampling scheme, different sampling pattern (mask) is used for each depth plane, whereas in fixed sampling we used same sampling pattern (mask) for all depth planes. 
The network was trained for variable sampling scheme only and both sampling schemes were used in test phase.
The variable sub-sampling patterns can be implemented using a software changes by randomly selecting the A/D converter,
of which procedure is similar to various compressive US researches, e.g. \cite{schretter2017ultrasound}.}

\subsection{Network training}

As for the target IQ data, we mainly used the IQ data from DAS beamforming results from the full RF data.
{We also use IQ data from an adaptive beamformer and deconvolution beamformer  \cite{7565583} to demonstrate that the proposed beamformer
can be trained to mimic various types of beamformers.}

The network was implemented with MatConvNet \cite{vedaldi2015matconvnet} in the MATLAB 2015b environment. Specifically, for network training, the parameters were estimated by minimizing the $l_2$ norm loss function using a stochastic gradient descent with a {$l_2$} regularization parameter of $10^{-4}$. The learning rate started from $10^{-4}$ and {exponentially} decreased to $10^{-7}$ in $200$ epochs. The weights were initialized using Gaussian random distribution with the Xavier method \cite{glorot2010understanding}. 

It is noteworthy to highlight this important aspect of our DeepBF model that {for each target beamformer e.g., DAS, MVBF or deconvolution}, a single (one time trained) CNN model is used for all data types and sub-sampling rates.

\subsection{Comparison methods}
For the evaluation purpose, we compared our proposed DeepBF method with standard DAS and MV beamformers. 
In DAS, the beamforming step is a simple {weighted} sum.
Specifically,  for DAS formulation in \eqref{eq:DAS}, $J$ is varied from 64 to 4 depending on the sub-sampling ratios so that data from $J$ active receivers is added to generate beamformed output. 
{
For the adaptive beamforming case, $\Rb_{l,n}$ must be estimated with a limited amount of data. A widely used method for the estimation of $\Rb_{l,n}$ is spatial smoothing (or subaperture averaging) \cite{AdaptiveBFMUS}, in which  the sample sub-aperture covariance matrix is calculated  by averaging covariance matrices of $K$ consecutive channels in the $J$ receiving channels.
Here we use $K=16$. 
Then, the weight for the minimum variance beamformer is calculated using the sub-aperture covariance estimate,
after which the final beamforming result is obtained
 by averaging the contribution from the adaptive beamforming
results from each subaperture array.
}

\subsection{Performance metrics}

 To quantitatively show the advantages of the proposed deep learning method, we used the {contrast-recovery} (CR), contrast-to-noise ratio (CNR) \cite{BiomedicalImageAnalysis}, generalized CNR (GCNR) \cite{GCNR_Paper}, peak-signal-to-noise ratio (PSNR), structure similarity (SSIM) \cite{1284395} and the reconstruction time.  
 
  The contrast is measured for the background ($B$) and anechoic structure ($aS$) in the image, and quantified in terms of
   CR and CNR:
 \begin{equation}
 {\hbox{CR}}(B,aS) = |\mu_{B}-\mu_{aS}|
 \end{equation}
 \begin{equation}
 {\hbox{CNR}}(B,aS) = \frac{|\mu_{B}-\mu_{aS}|}{\sqrt{\sigma^2_{B} + \sigma^2_{aS}}},
 \end{equation}
  where $\mu_{B}$, $\mu_{aS}$, and $\sigma_{B}$, $\sigma_{aS}$ are the local means, and the standard deviations of the background ($B$) and anechoic structure ($aS$) \cite{BiomedicalImageAnalysis}, respectively. 
  Another improved measure for the contrast-to-noise-ratio called generalized-CNR (GCNR) was recently proposed  \cite{GCNR_Paper}. The GCNR compare the overlap between the intensity distributions of two regions.  The GCNR is defined as
 \begin{equation}
 {\hbox{GCNR}}(B,aS) = 1- \int \min \{p_{B} (x), p_{aS} (x) \} dx,
 \end{equation}
 where $x$ is the pixel intensity,  and $p_{B}$ and $p_{aS}$ are the probability distributions of the background ($B$) and anechoic structure ($aS$), respectively. If both distribution are completely independent,  then GCNR will be equals to one, whereas, if they completely overlap then GCNR will be zero \cite{GCNR_Paper}.
 The GCNR measure is difficult to tweak, so we believe that GCNR is an objective performance metric. For CNR and GCNR calculations, we generated separate ROI masks for each image.

  The PSNR and SSIM index are calculated on reference ($v$) and Rx sub-sampled ($\tilde v$) images of common size  $n_1\times n_2$ as

\begin{equation}
{\hbox{PSNR}}(v,\tilde v) = 10 \ensuremath{\log_{10}} \left(\frac{R_{\max}^2}{\|v-\tilde v\|_F^2}\right),
\end{equation}
where $\|\cdot\|_F$  denotes the Frobenius norm and $R_{\max}=2^{(\#bits\ per\ pixel)}-1$ is the dynamic range of pixel values (in our experiments this is equal to $255$),
and
\begin{equation}
{\hbox{SSIM}}(v,\tilde v) = \frac{(2\mu_{v}\mu_{\tilde v}+c_{1})(2\sigma_{v,\tilde v} +c_{2})}{(\mu^{2}_{v}+\mu^{2}_{\tilde v}+c_{1})(\sigma^{2}_{v}+\sigma^{2}_{\tilde v} +c_{2})},
\end{equation}
where  $\mu_{v}$, $\mu_{\tilde v}$, $\sigma_{v}$, $\sigma_{\tilde v}$, and $\sigma_{v,\tilde v}$ are the local means, standard deviations, and across-covariance for images $v$ and $\tilde v$ calculated for a radius of $50$ units.  The default values of $c_{1}=(k_{1}R_{max})^{2}$, $c_{2}=(k_{2}R_{max})^{2}$, $k_{1}=0.01$ and $k_{1}=0.03$.

\begin{table*}[!hbt]
	\centering
	\caption{Performance statistics on \textit{in vivo} data for random sampling pattern}
	\label{tbl:results_vSTATS_invivo}
	\resizebox{\textwidth}{!}{
		\Huge
		\begin{tabular}{c|cccc|cccc|cccc|cccc|cccc}
			\hline
			{sub-sampling} & \multicolumn{4}{c|}{{CR (dB)}} & \multicolumn{4}{c|}{{CNR}} & \multicolumn{4}{c|}{{GCNR}} & \multicolumn{4}{c|}{{PSNR (dB)}} & \multicolumn{4}{c}{{SSIM}}  \\
			{factor} & \textit{DAS} & \textit{MVBF} & \textit{DeepBF} & \textit{DeepMVBF} & \textit{DAS} & \textit{MVBF} & \textit{DeepBF} & \textit{DeepMVBF} & \textit{DAS} & \textit{MVBF} & \textit{DeepBF} & \textit{DeepMVBF} & \textit{DAS} & \textit{MVBF} & \textit{DeepBF} & \textit{DeepMVBF} & \textit{DAS} & \textit{MVBF} & \textit{DeepBF} & \textit{DeepMVBF}  \\ \hline\hline
			1 & 12.37 & 12.39 & 13.25 & 13.25 & 1.38 & 1.38 & 1.45 & 1.45 & 0.64 & 0.64 & 0.66 & 0.66 & $\infty$ & $\infty$ & $\infty$ & $\infty$ & 1 & 1 & 1 & 1 \\
			2 & 10.55 & 10.62 & 13.05 & 13.26 & 1.33 & 1.33 & 1.47 & 1.47 & 0.63 & 0.63 & 0.66 & 0.66 & 24.59 & 24.63 & 27.38 & 27.71 & 0.89 & 0.89 & 0.95 & 0.95 \\
			2.7 & 10.06 & 10.16 & 12.63 & 13.29 & 1.30 & 1.30 & 1.44 & 1.48 & 0.62 & 0.62 & 0.66 & 0.67 & 23.15 & 23.17 & 25.54 & 26.08 & 0.86 & 0.86 & 0.92 & 0.93 \\
			4 & 9.54 & 9.65 & 11.80 & 13.14 & 1.25 & 1.25 & 1.38 & 1.47 & 0.60 & 0.60 & 0.65 & 0.66 & 21.68 & 21.75 & 23.55 & 24.29 & 0.81 & 0.81 & 0.87 & 0.89 \\
			8 & 9.05 & 9.21 & 10.47 & 12.17 & 1.18 & 1.19 & 1.26 & 1.37 & 0.58 & 0.58 & 0.61 & 0.64 & 19.99 & 20.02 & 21.03 & 21.96 & 0.74 & 0.74 & 0.78 & 0.82  \\
			16 & 8.98 & 9.11 & 9.73 & 10.99 & 1.12 & 1.11 & 1.17 & 1.25 & 0.56 & 0.56 & 0.58 & 0.61 & 18.64 & 18.68 & 19.22 & 20.24 & 0.67 & 0.67 & 0.69 & 0.74 \\  \hline
		\end{tabular}
	}
	\vspace*{-0.3cm}
\end{table*}

\begin{figure}[!h]
	\centerline{\includegraphics[width=7.cm]{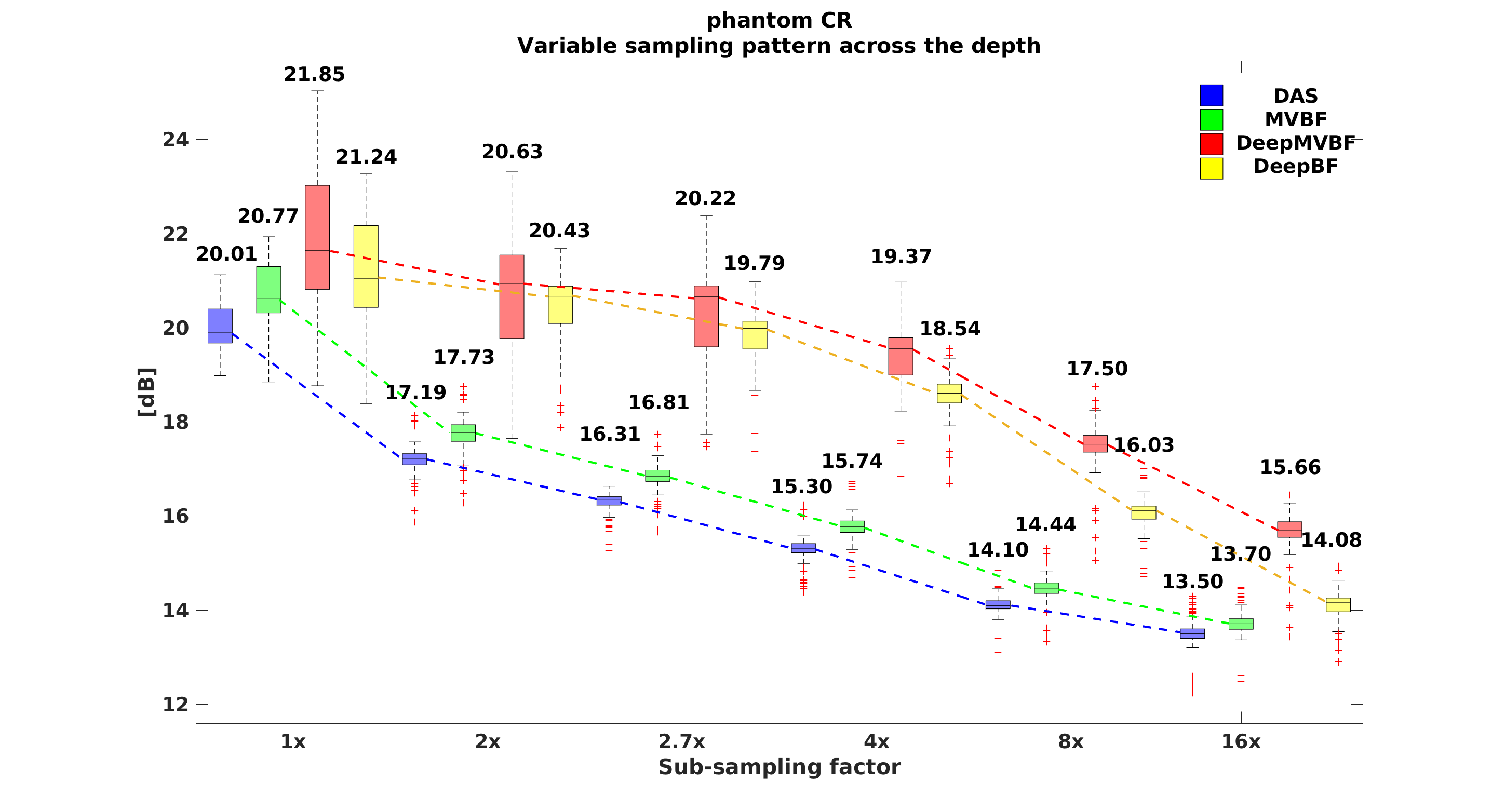}}
	\centerline{\includegraphics[width=7.cm]{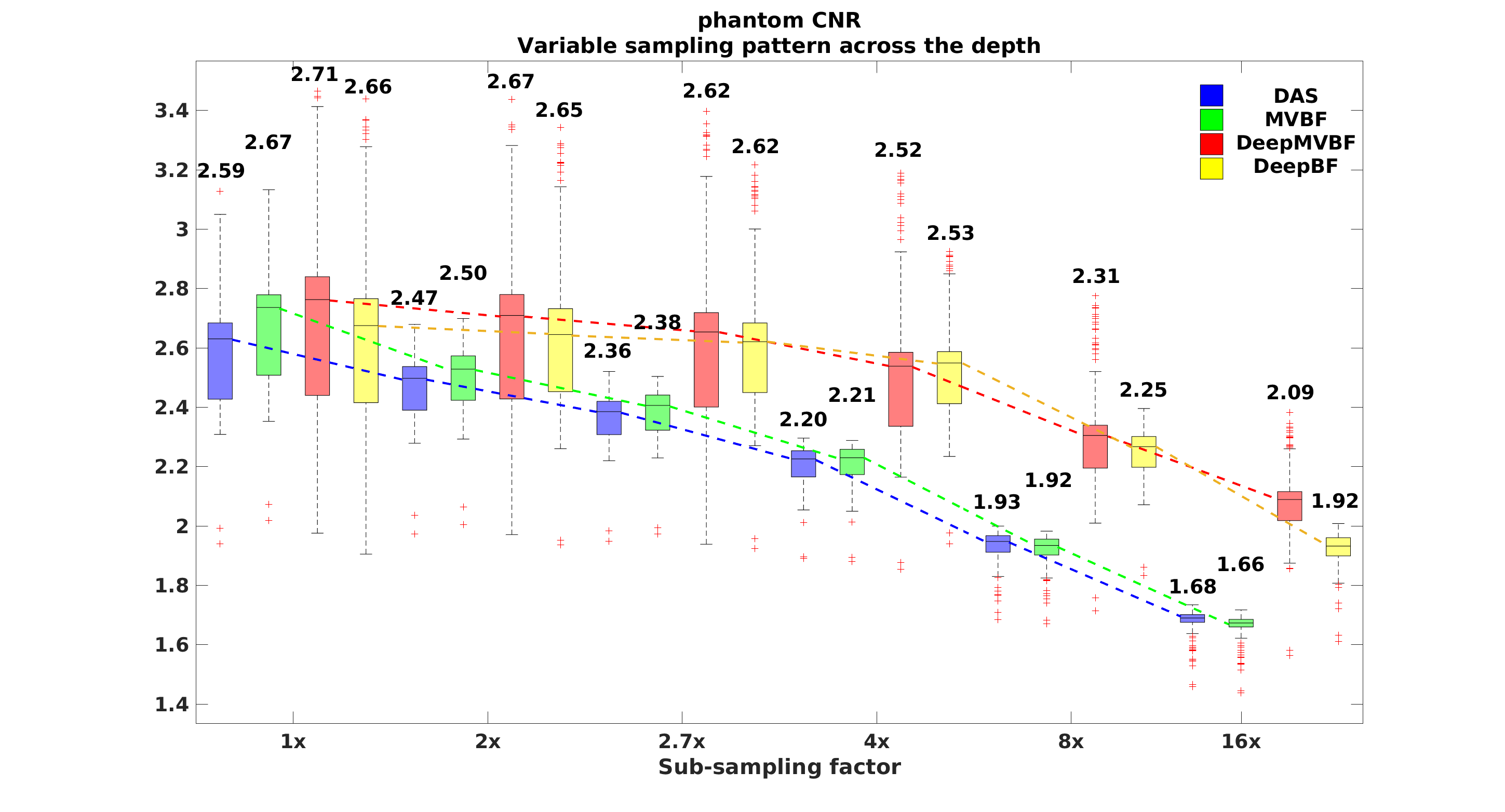}}
	\centerline{\includegraphics[width=7.cm]{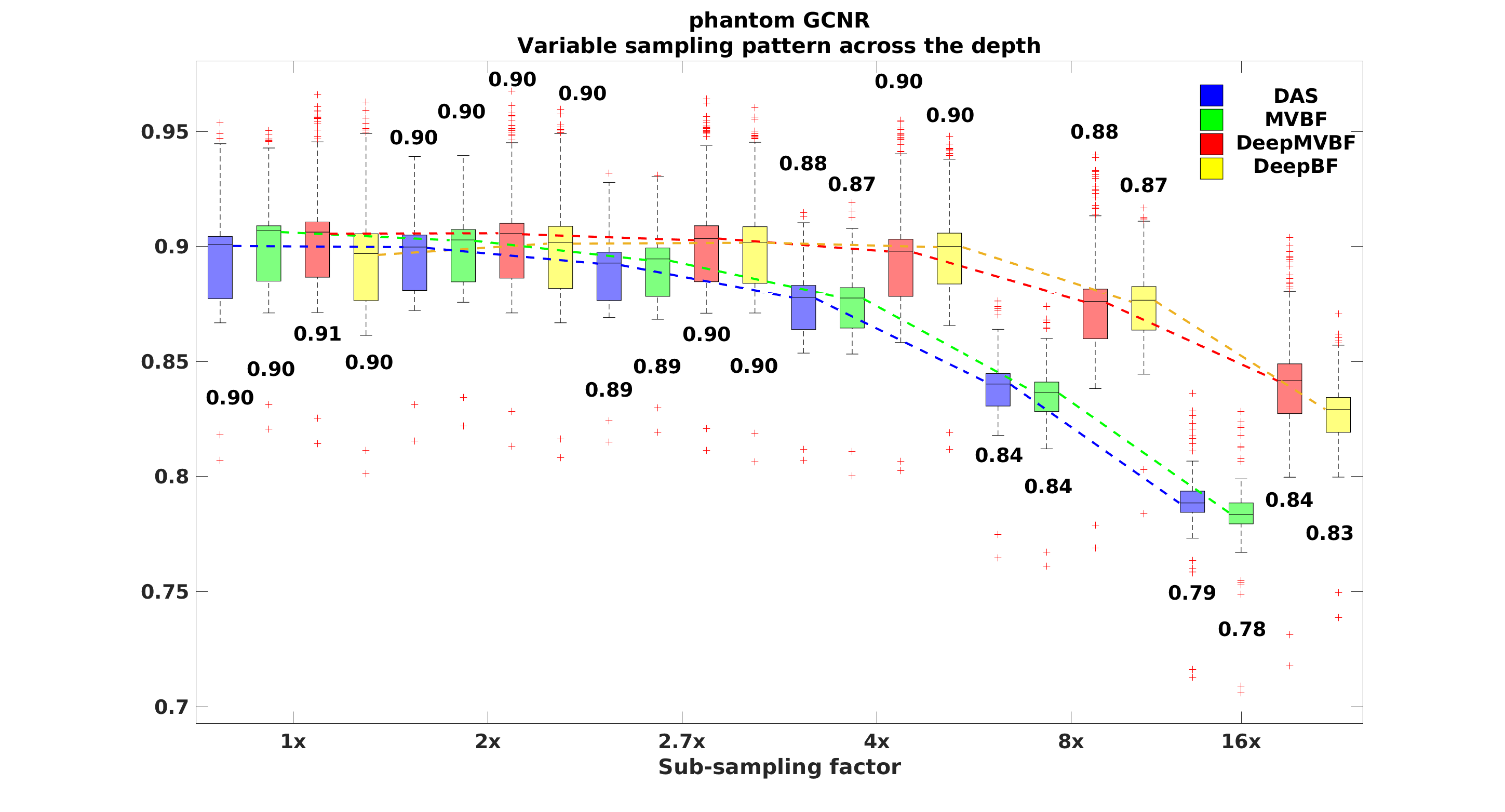}}
	\centerline{\includegraphics[width=7.cm]{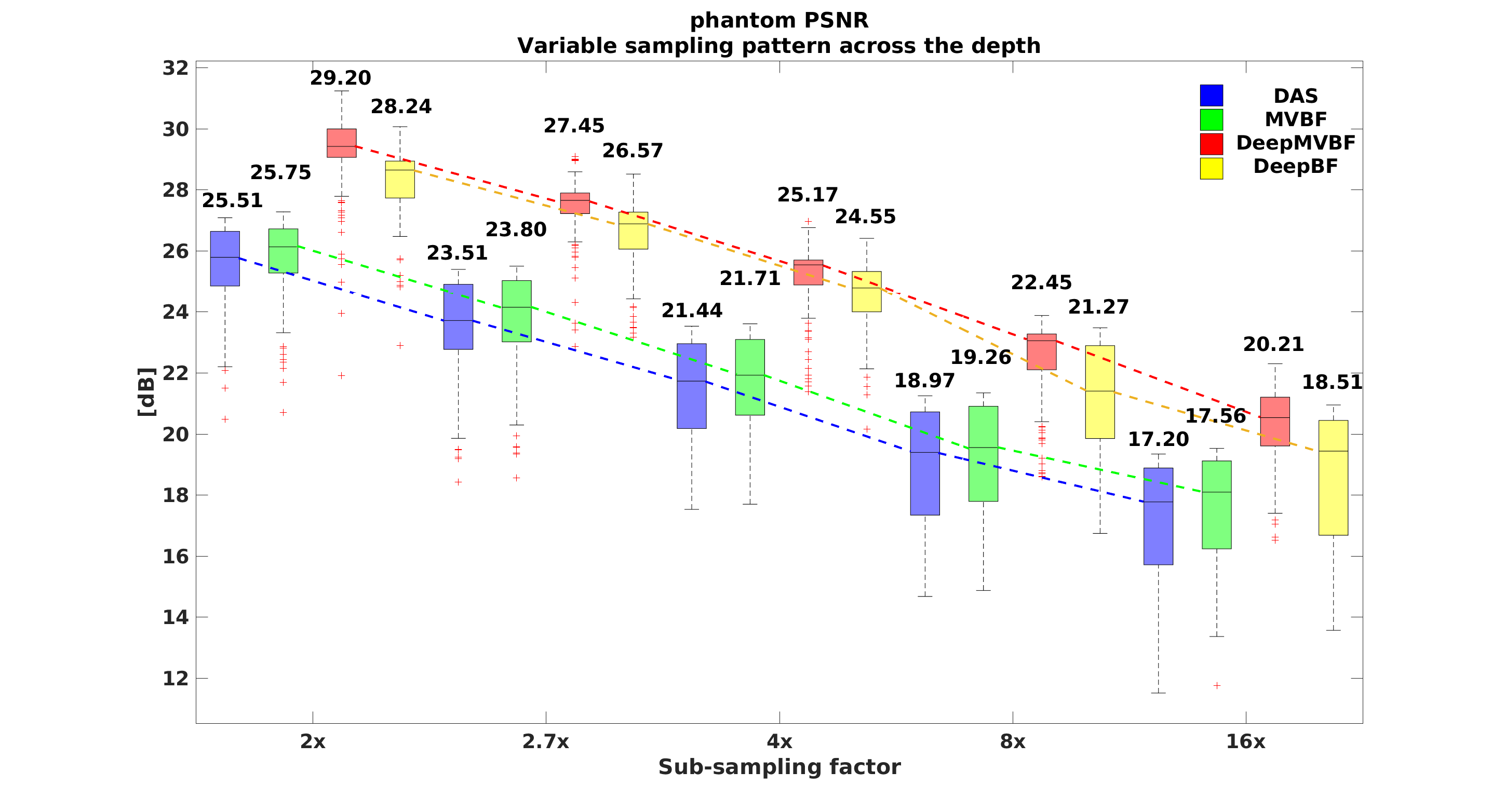}}
	\centerline{\includegraphics[width=7.cm]{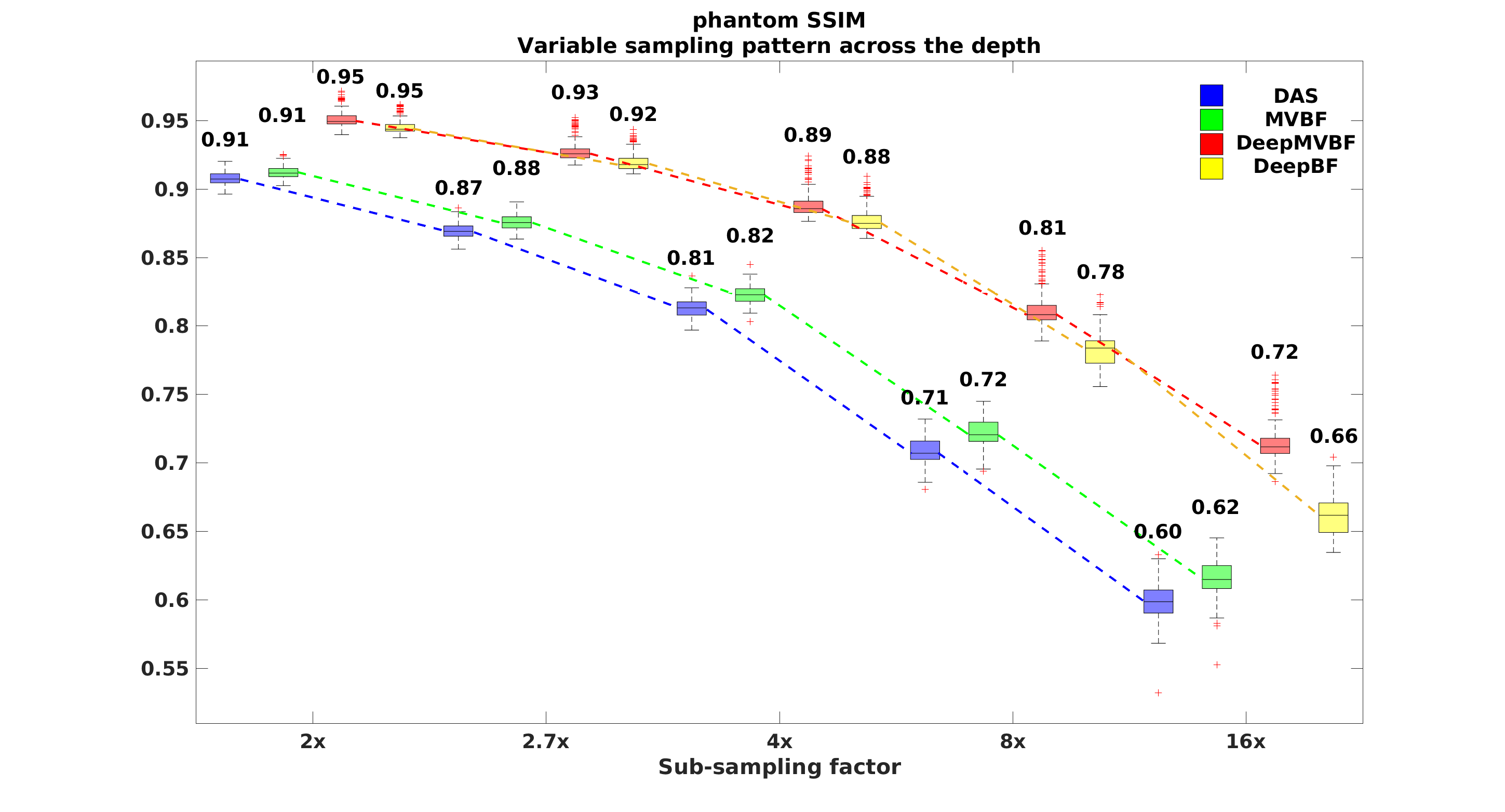}}
\vspace*{-0.3cm}
	\caption{\footnotesize Quantitative comparison using phantom data with respect to different subsampling {rates}: ({first row}) CR value distribution, ({second row}) CNR value distribution, ({third row}) GCNR value distribution, ({fourth row}) PSNR value distribution, ({fifth row}) SSIM value distribution }
	\label{fig:results_STATS_phantom}
\end{figure}


\section{Experimental Results}
\label{sec:results}

In this section we present extensive comparison results of our method with DAS and the minimum variance beamformer (MVBF) 
for various acquisition scenarios.
%
{Our DeepBF was first trained with DAS data obtained from full RF data, and
we compare our results with those by DAS and the minimum variance beamformer (MVBF).
We also trained our neural network using the MVBF from full RF data. With a slight abuse of terminology,
this is often  referred to as DeepMVBF,
although we use the term DeepBF to refer general deep neural network-based beamformers regardless of specific target data for  training.
}

\subsection{Quantitative comparison}

We  compared the CR, CNR, GCNR, PSNR, and SSIM distributions of reconstructed B-mode images of \textit{in vivo} and phantom test datasets. 
Table~\ref{tbl:results_vSTATS_invivo}  shows the comparison of DAS, MVBF, proposed DeepBF, {and DeepMVBF methods} on $360$ \textit{in vivo} test frames for random sub-sampling scheme. In terms of CR, CNR and GCNR,  the overall performance of DAS and MVBF were relatively similar. However, the results by the proposed methods 
 are significantly superior to those of DAS and MVBF at various subsampling factors.
To investigate the performance degradation with respect to the subsampling,
in Table~\ref{tbl:results_vSTATS_invivo} we also show the PSNR and SSIM values with respect to the results of the full scan. 
Again, the performance degradation in terms of PSNR and SSIM was much less by the proposed DeepBF {and DeepMVBF}. 


In Fig.~\ref{fig:results_STATS_phantom}, we provide  distribution plots for  various performance measures for phantom test dataset. In fully sampled case, our DeepBF shows overall gain of around $1.23$ dB in CR, and $0.07$ units improvement in CNR compared to DAS. In sub-sampling cases, unlike DAS and MVBF in which the performance is highly sensitive to the rate of sub-sampling, the proposed DeepBF shows consistent GCNR performance even at $4\times$ reduced sampling rate. This can be easily seen in Fig.~\ref{fig:results_STATS_phantom} (third row), where the average value of GCNR remain constant at $0.90$ units for $1\times$ to $4\times$ sampling factors and only drop by $0.03$ and $0.04$ units at $8\times$ and $16\times$ sampling factors respectively.  {The performance of DeepBF and DeepMVBF were similar, although DeepMVBF has slight better CR values.}

The CR, CNR, and GCNR are measure for local regions, whereas the PSNR and SSIM are global metric. 
{To calculate the PSNR and SSIM, images generated using $64$ Rx-channels were considered as reference images for all algorithms.} As shown in Fig.~\ref{fig:results_STATS_phantom} the proposed methods show significantly higher PSNR and SSIM values, confirming that
 the proposed methods successfully recover actual structural detail in sub-sampled images.

\subsection{Qualitative Comparison}

{To verify the performance improvement in terms of visual quality,  here we provide
representative reconstruction results. Due to the similarity between the DeepBF and DeepMVBF, we only
provide the results by DeepBF in this section.}

\subsubsection{Full RF data cases}

\begin{figure}[!hbt]
\centering
\centerline{\epsfig{figure=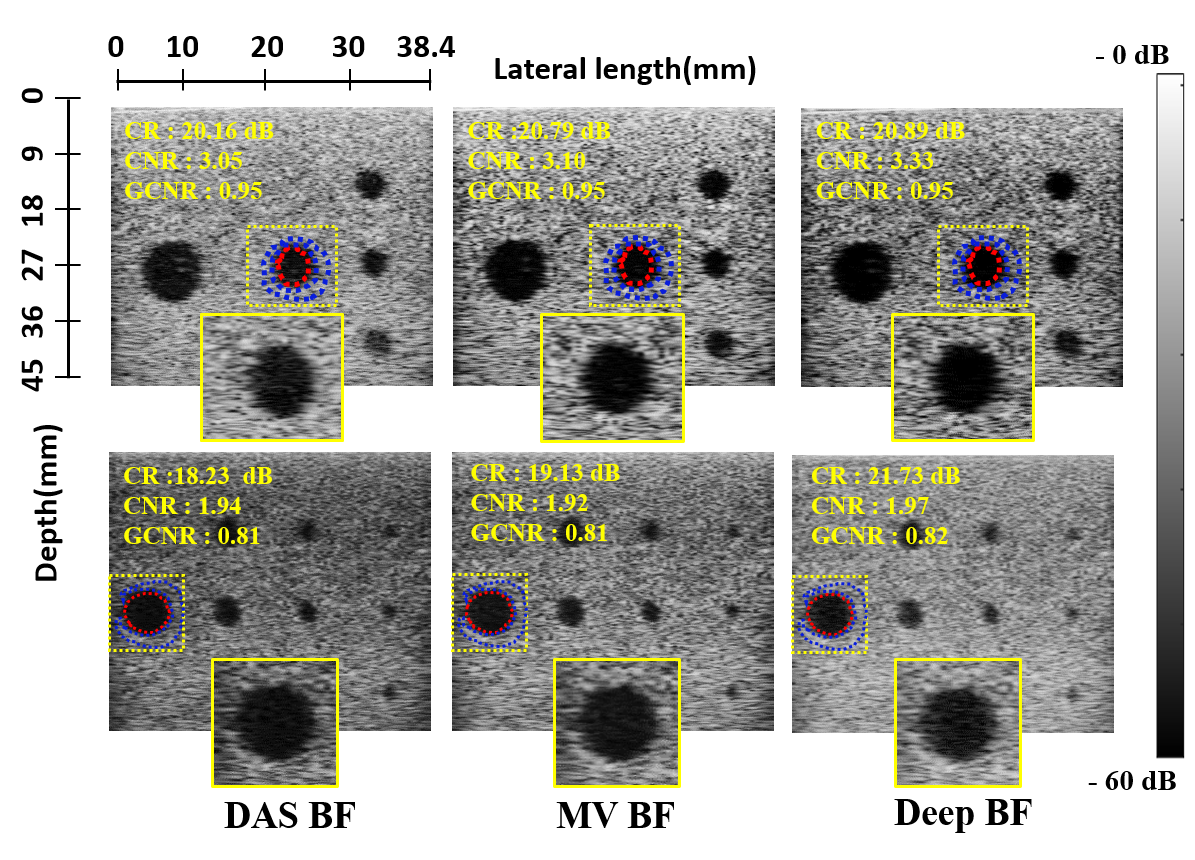, width=9cm}}   
\caption{Reconstruction results of standard DAS BF, MV BF and the proposed DeepBF for fully sampled phantom scans data using $8.5$ MHz center frequency.}
\label{fig:results_view_phantom_full}	
\end{figure}

 In {Fig}.~\ref{fig:results_view_phantom_full} we compared two phantom examples scanned using $8.5$ MHz center frequency.  In phantom test dataset, the proposed DeepBF achieves comparable or even 
 better performance compared to DAS and MVBF methods.  From the figures 
 we found that  the visual quality of DeepBF reconstruction,  especially around anechoic regions, is  comparable or better than  that of MVBF method, which is better than DAS beamformer.  
Quantitatively,  CR, CNR, and GCNR values of Deep BF were slightly improved compared to the existing methods.
{It is also noteworthy to point-out that even though the proposed DeepBF was only trained on \textit{in vivo} carotid/thyroid data scanned with $8.5$MHz operating frequency, its performance in very diverse test scenarios is still remarkable, which clearly shows the generalization power of the proposed method. }

\begin{figure}[!hbt]
	\centerline{\epsfig{figure=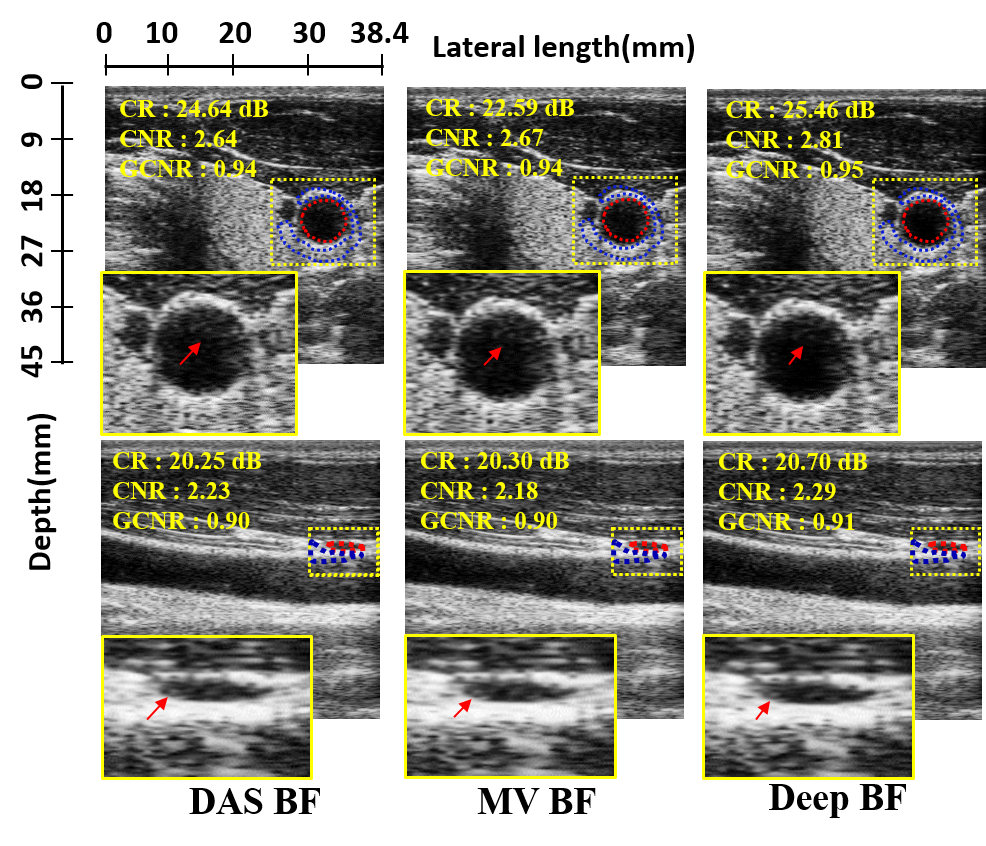, width=9cm}}   
	\caption{\footnotesize Reconstruction results of standard DAS-BF, MVBF and the proposed DeepBF for fully sampled in vivo scans from Carotid region.
	All the figures are illustrated at the same window level ($0\sim -60$dB for the beamformed images).}
	\label{fig:results_view_invivo_full}	
\end{figure}

  To further validate the performance gain on fully-sampled data,
  in Fig.~\ref{fig:results_view_invivo_full} we showed the results of two \textit{in vivo} examples for fully-sampled data.  The images are generated using standard DAS, MVBF and the proposed DeepBF method.   In {Fig}.~\ref{fig:results_view_invivo_full}, it can be easily seen that our method  provides visual 
  quality comparable to DAS and MVBF methods. Interestingly, it is remarkable that the CR, CNR and GCNR values are improved by the DeepBF.
  To investigate the origin of the quantitative improvement, we showed the magnified views as inset in {Fig}.~\ref{fig:results_view_invivo_full}.
  With a careful look, we can see that there are several artifacts around the wall of anechoic regions in DAS and MVBF methods, which can be confused
  with structure. On the other hand, those artifacts are not visible in DeepBF, which {makes} the 
  the visual quality of the US images and quantitative values  higher compared to DAS and MVBF methods.   

\begin{figure}[!t]
		\centerline{\epsfig{figure=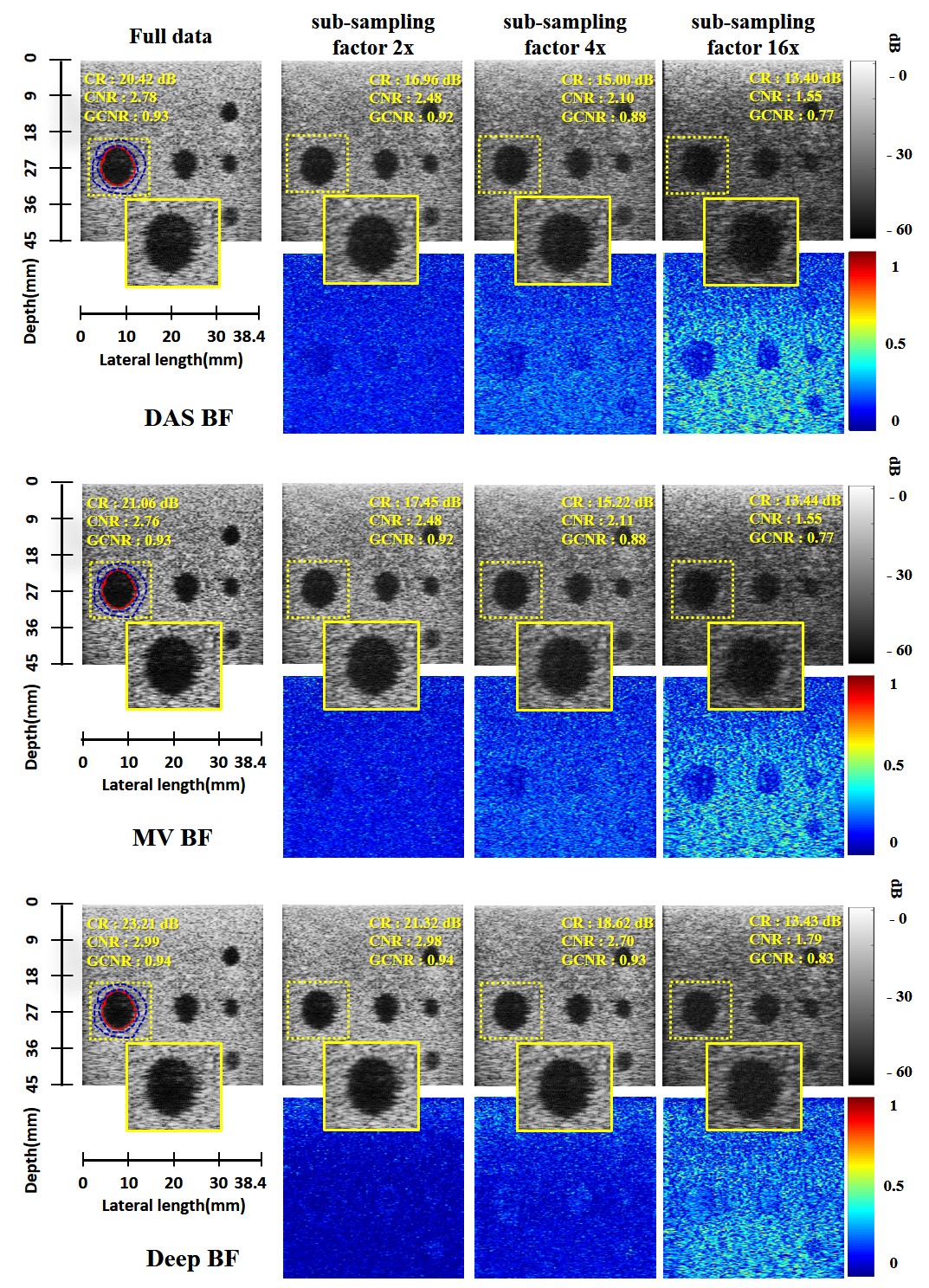, width=9cm}}   
	\caption{\footnotesize Comparison of various beamformers for phantom data with random sub-sampling patterns.  The difference image is normalized to $0\sim 1$ scale for each image.}
	\label{fig:random_phantom}	
\end{figure}

\begin{figure*}[!t]
	\centerline{\epsfig{figure=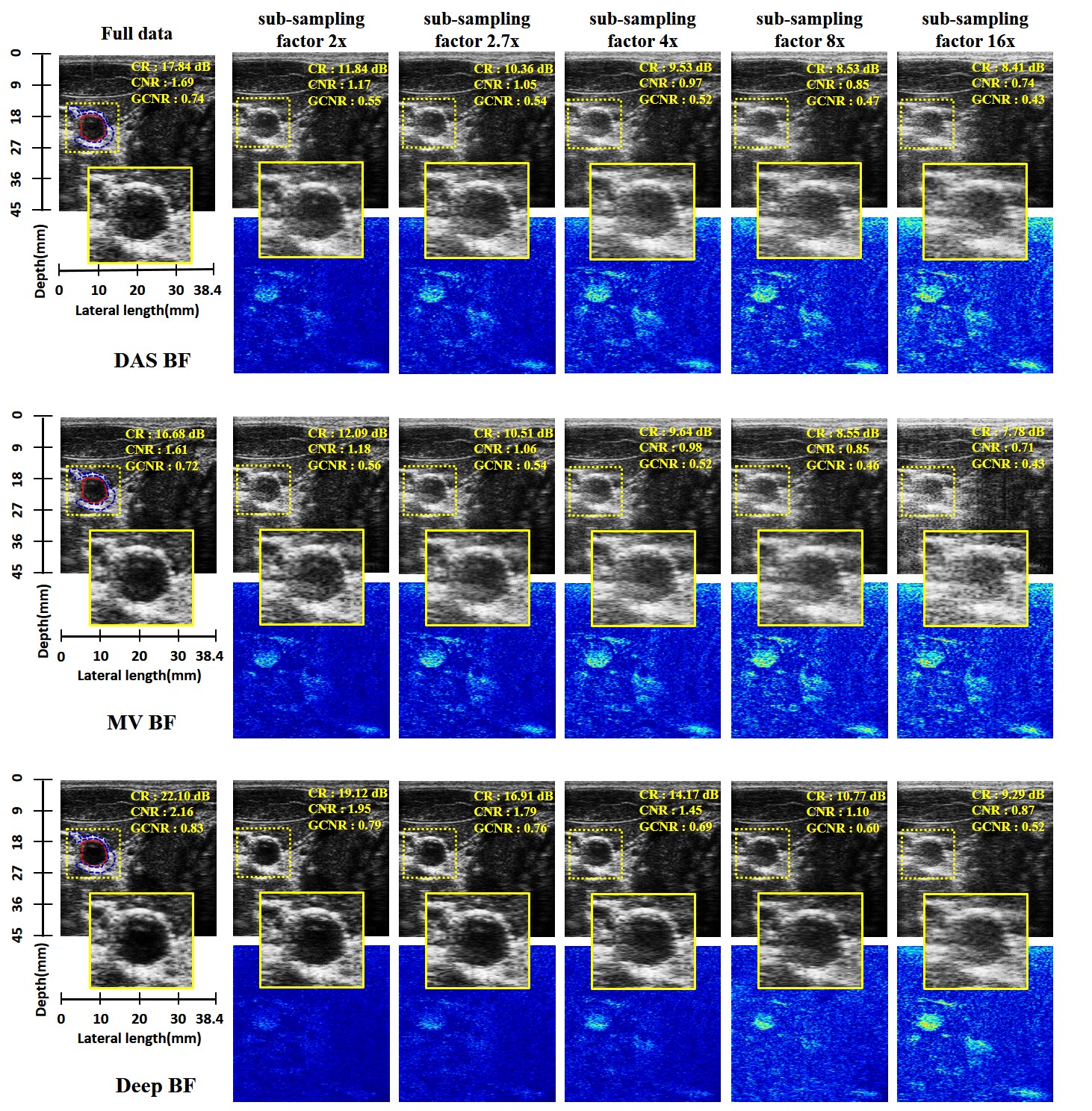, width=18cm}}   
	\caption{\footnotesize Comparison of various beamformers for in vivo data with random sub-sampling patterns. The images are shown in the same window
	level ($0\sim -60$dB for the beamformed images, and  the difference image is normalized to $0\sim 1$ scale for each image.)}
	\label{fig:random_invivo}	
\end{figure*} 

\subsubsection{Compressive beamforming}

%
  
   Fig.~\ref{fig:random_phantom}
   show the results of a phantom example for $32$, $16$, and $4$ Rx-channels down-sampling schemes on random sampling scheme.  Since 64 channels are used as a full sampled data, this corresponds to the full data as well as subsampled data with $2\times, 4\times$ and $16\times$ sub-sampling factors.  The images are generated using the proposed DeepBF, MVBF and the standard DAS beamformer methods.   Our method significantly improves the visual quality of the US images by estimating the correct structural details and eliminating artifacts caused by sub-sampling.  The residual of fully-sampled and sub-sampled images are shown in pseudo colors on normalized scale. From the normalized difference images it can be easily seen that   DeepBF produces uniformly distributed noise-like errors across various subsampling ratios, whereas  DAS and MVBF produce structure-dependent errors that can reduce the image contrast.
   Note that the training data consist of only \textit{in vivo} {carotid/thyroid} scans; but relative improvement in diverse test scenarios is nearly the same for both \textit{in vivo} and phantom cases. This further confirms the generalization power of the proposed method. 



{Fig}.~\ref{fig:random_invivo}  illustrates  representative examples of \textit{in vivo} data at $2\sim 16$x acceleration. By harnessing the spatio-temporal (multi-depth and multi-line) learning, the proposed CNN-based beam-former successfully reconstructs the images with good quality in all down-sampling schemes.  From residual images it can be seen that in contrast to DAS and MVBF, 
 the proposed DeepBF maintains good visual quality at even at highest sub-sampling rate.  Unlike DAS and MVBF, DeepBF preserves the original structural details as well as the contrast of the sub-sampled data much closer to the fully-sampled image. 

\subsection{Computational time}
One big advantage of ultrasound image modality is {its} run-time imaging capability, {which allows for fast reconstruction times}. Although training required $40$ hours for $200$ epochs using MATLAB, once training was completed, the reconstruction time for the proposed deep learning method is not very long. The average reconstruction time for each depth planes is around $4.8$ (milliseconds), which could be easily reduced by optimized implementation and reconstruction of multiple depth {plane} in parallel.

\begin{figure}[!hbt]
	\centerline{\includegraphics[width=9cm]{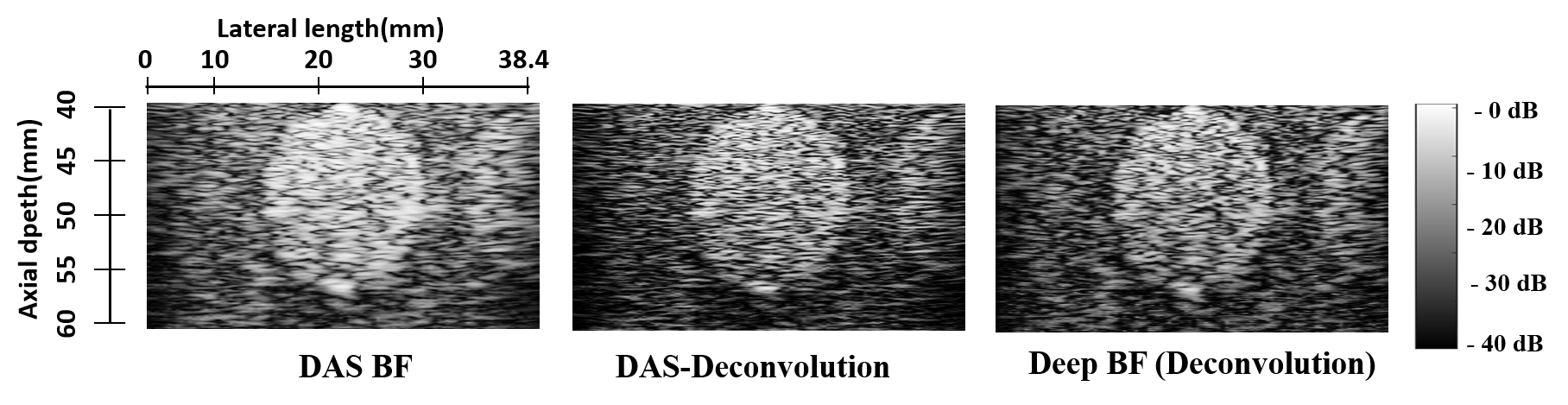}}
		\vspace*{-0.3cm}	
	\caption{Comparison of DAS, deconvolution of DAS, and proposed DeepBF with deconvoluted DAS as label using phantom anechoic cyst of $15$ mm diameter at $40$ mm depth on fully sampled RF-data.}
	\label{fig:results_Deconv}
\end{figure}

\begin{figure*}[!hbt]
	\centerline{\epsfig{figure=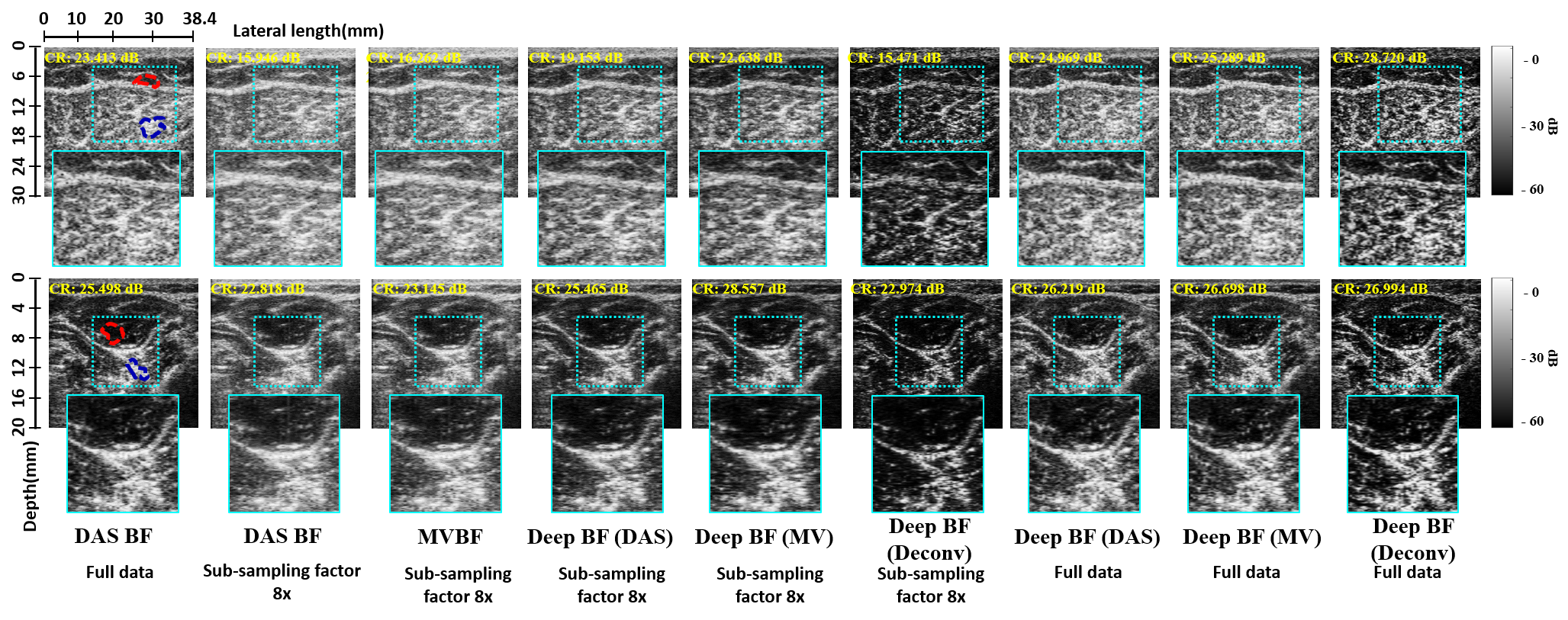, width=18cm}}   
		\vspace*{-0.3cm}
	\caption{\footnotesize Reconstruction results from (left) forearm muscle (right) calf muscle using standard DAS, MVBF, and Deep BFs at various random subsampling factors.
	Here, various form of Deep BF were trained  using DAS, MVBF, and deconvoluted DAS as target data from fully sampled in vivo scans.
		All the figures are illustrated at the same window level ($0\sim -60$dB for the beamformed images).}
	\label{fig:Different_Body_Parts}	
\end{figure*}

\section{Discussion}
\label{sec:discussion}

\subsection{Expressivity and Generalization}

{To confirm that the proposed neural network learns various target data,
 we also trained our model using deconvolution of DAS results. The training was performed using fine-tuning method using target data generated by a deconvolution method using sparse representation as described in \cite{7565583}. In Fig~\ref{fig:results_Deconv}, we compared the reconstruction results by DAS, deconvolution of DAS, and the proposed DeepBF trained with deconvoluted DAS as target, using phantom anechoic cyst of $15$ mm diameter at $40$ mm depth on fully sampled RF-data. From the results it can be easily seen that the proposed method can successfully learn to mimic the deconvolution method.}

\begin{figure*}[!t]
	\centerline{\epsfig{figure=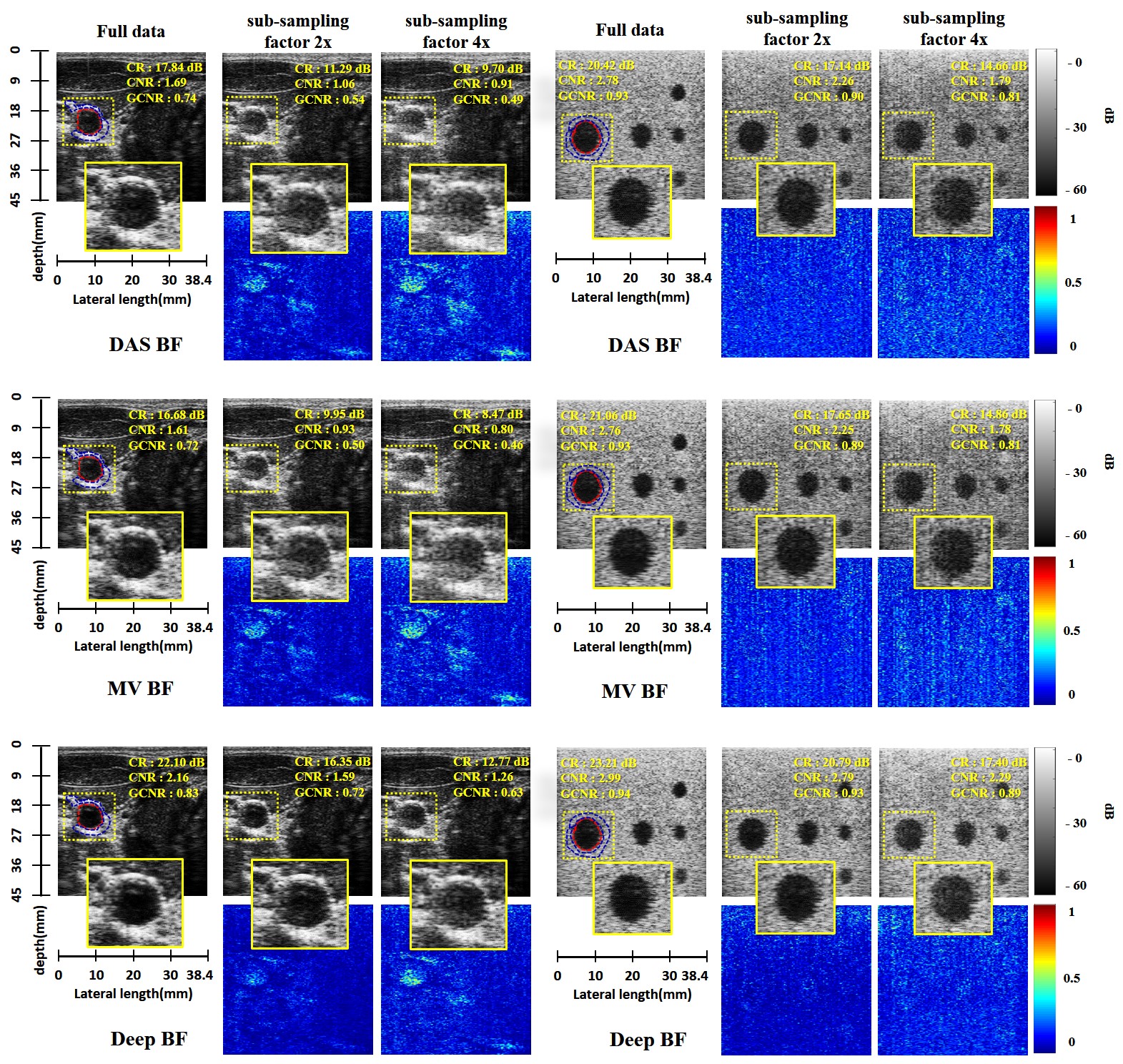, width=18cm}}   
	\caption{\footnotesize Comparison of various beamformers for phantom data using uniform sub-sampling patterns.  The difference image is normalized to $0\sim 1$ scale for each image.}
	\label{fig:uniform_invivo}	
\end{figure*}

\begin{figure}[!hbt]
	\centerline{\includegraphics[width=9cm]{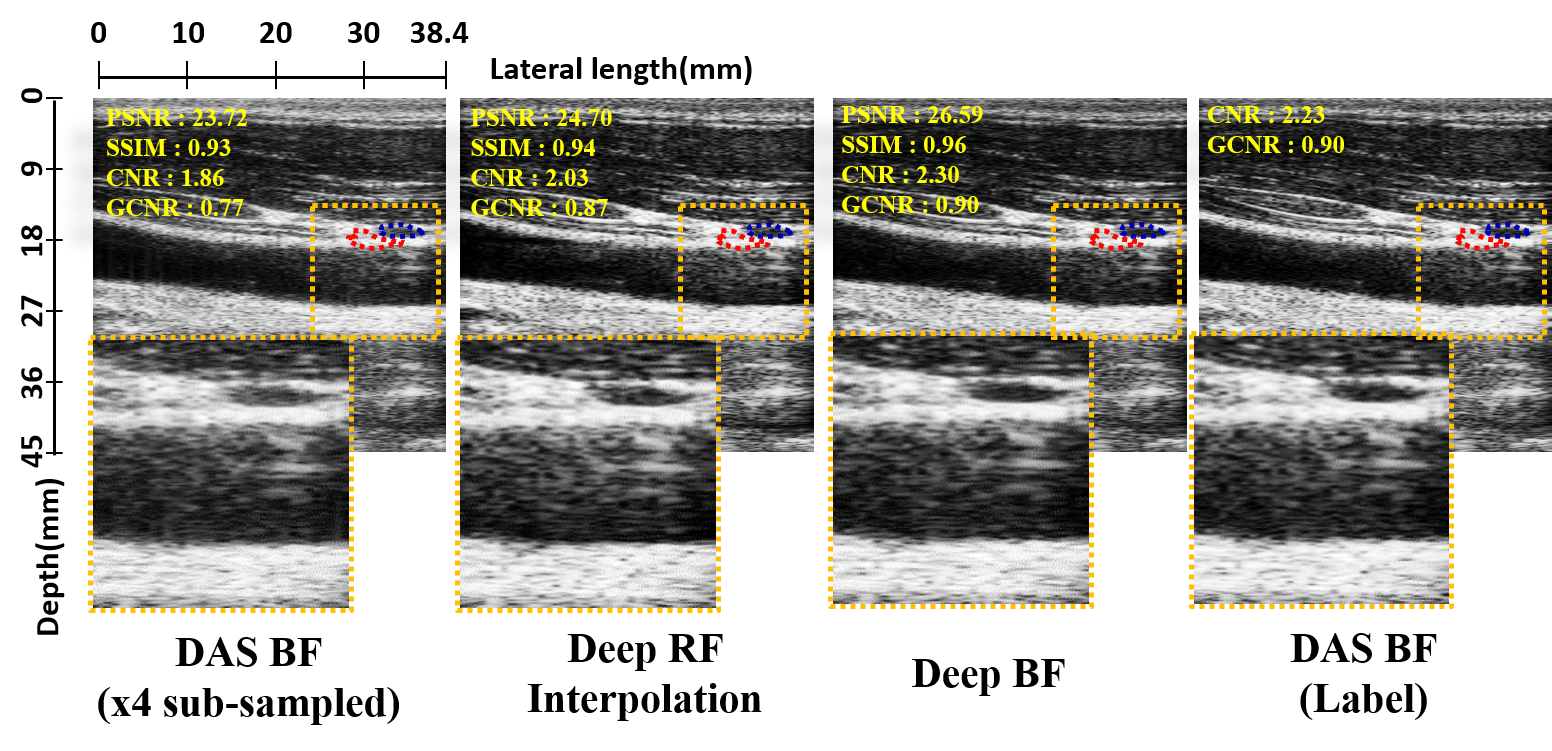}}	
	\caption{Reconstruction results of standard DAS beam-former, Deep RF Interpolation \cite{yoon2018efficient} and the proposed DeepBF for 4$\times$ sub-sampled in vivo data. The window levels are same for all images ($0\sim -60$dB).}
	\label{fig:results_Othermethods}
\end{figure}

\begin{figure*}[!hbt]
	\centerline{\includegraphics[width=18cm]{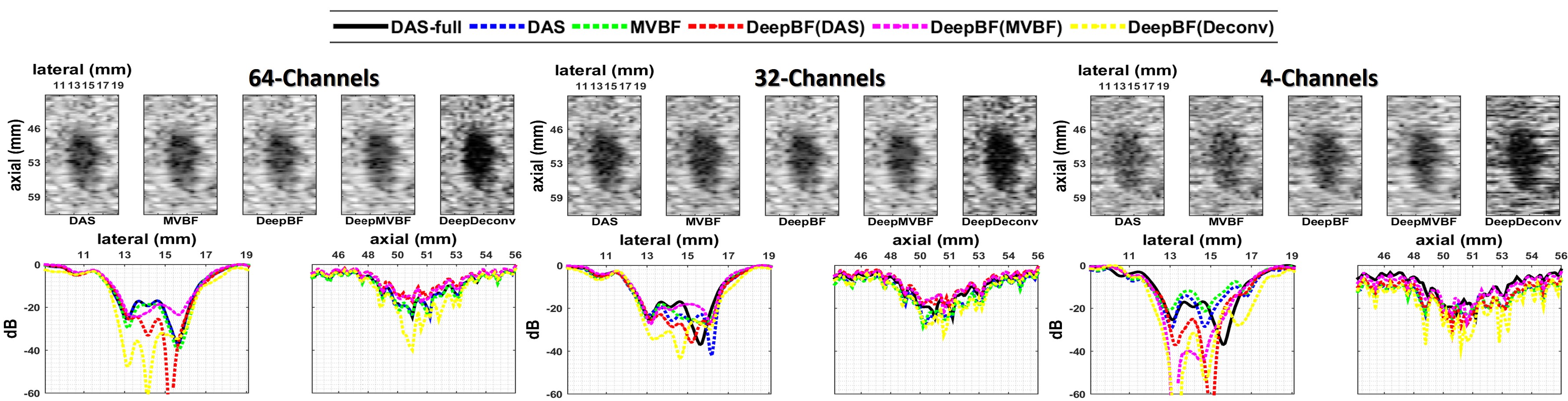}}	
	\caption{Lateral and axial profiles across the depth through the center of the phantom $6$ mm diameter anechoic cyst at $48$ mm depth using DAS and various form of DeepBFs using random sampling patterns. Images are shown with a $60$ dB dynamic range.}
	\label{fig:results_SP_vLin}
\end{figure*}

To further compare the effect of various choices of target data and its generalization power, we evaluated the network trained with
different label data  by testing on completely different datasets. 
Specifically, {we trained our neural network models using fully sampled carotid/thyroid} DAS images, MVBF images, and deconvolution  results.
Then, new test dataset were acquired from forearm and calf muscles using $10$MHz carrier frequency.  In Fig~\ref{fig:Different_Body_Parts}, it is evident that the proposed DeepBF method can successfully process RF-data from different anatomical region and operating frequencies. 
Among the various choices of the target data, the performance using  the deconvoluted DAS target was best with less clutters,
which was followed by DeepBF trained using MVBF targets and DAS targets. It is also noteworthy to point out that neural network trained with DAS targets still provides better
results than the standard DAS and MVBF for x8 subsampling cases. In particular, on average DAS and MVBF showed $21.11$ dB and $21.38$ dB CR which is $2.06$ dB and $1.79$ dB less than the DeepBF respectively.  Moreover, the performance of DeepBF and DeepMVBF were again somewhat similar.

\subsection{Dependency on the Sampling Patterns}

 {So far we showed that a considerable reduction in data rate can be achieved by applying DeepBF method, leading to a sub-Nyquist sampling rate, which uses only a portion of the bandwidth of the ultrasound signals to reconstruct the high quality image. 
 Another application of sub-sampled US is the reduction of active Rx elements. For this purpose we also evaluated our model using uniform sub-sampling scheme  to confirm whether relative performance gain with uniform sub-sampling is also similar to the random sub-sampling. 
  In Table~\ref{tbl:results_sSTATS_invivo} we compared the performance of proposed DeepBF method with DAS and MVBF on \textit{in-vivo} data for uniform sub-sampling pattern. We can see in uniform sub-sampling case DeepBF achieves relatively similar performance gain as in the case of random sub-sampling.  In Fig~\ref{fig:uniform_invivo} two visual examples are shown for \textit{in-vivo} and phantom {data}, and we found that the proposed method successfully {reconstructs} the uniformly sub-sampled RF-data in both cases with equal error rates. }

\subsection{Ablation Studies}

 We also compared our method with Deep RF interpolation method \cite{yoon2018efficient}. Again, the proposed method also outperform the Deep RF interpolation method \cite{yoon2018efficient}. 
  Fig~\ref{fig:results_Othermethods} shows reconstruction results  of various methods at 4$\times$ subsampling rate, which is compared with the full data reconstruction. The contrast of the DeepBF, especially at anechoic regions, are very close to the full sampled case, whereas the other methods generates artifacts like patterns.
Quantitatively,  for 4$\times$ sub-sampled in vivo test dataset, the Deep RF interpolation \cite{yoon2018efficient} achieves CNR, GCNR, PSNR, and SSIM values of $1.31$, $0.63$ units, $22.15$ dB and $0.82$ units, which are $0.07$, $0.02$ units, $1.4$ dB and $0.05$ units inferior to the proposed method respectively.
Here we would like to point out that, in \cite{yoon2018efficient}, deep learning approach was designed for interpolating missing RF data, which are later used as input for standard beamformer. On the other hand, the proposed method is a CNN-based beamforming pipeline, without requiring additional beamformer.  Consequently, this approach is much simpler and can be easily incorporated to replace the standard beamforming pipeline.

The proposed multi-line, multi-depth method is also compared with different design strategies which include (1) reconstruction of RF sum without Hilbert
transform  (RF-sum only), (2) reconstruction of IQ data after training on fixed sub-sampling ratios (fixed sampling), and reconstruction of IQ data using only single depth plane (single-depth). Specifically, Table \ref{tbl:results_selfcomp_phantom} compares the performance of different design choices on phantom data for random sub-sampling case.  The results clearly show that the proposed methods of data-driven learning to generate IQ data using training data from multiple sub-sampling rates and multiple depths provides the best quantitative values.
{
Especially at higher sampling rates the multi-depth method show high PSNR and SSIM measures. Although we just used 3 depth planes in this experiment, for further improvements the idea can be generalized to different number of depth planes.}

\subsection{Improved Lateral Resolution}

In Fig~\ref{fig:results_SP_vLin}, we compared lateral and axis profiles through the center of a phantom anechoic cyst using DAS, MVBF and proposed methods.  In particular, an anechoic cysts of $6$mm diameter scanned from the depth of $52$mm and B-mode images were obtained for full data as well as random sampling  with {$2\times$} and $16\times$ sub-sampling factors using DAS, MVBF and proposed DeepBF methods. From the figures it can be seen that under all sampling schemes, on the boundary of cysts the proposed method show sharp changes in the pixel intensity with respect to the lateral position in the image. 
Although the axial profile shows similar trend to DAS for all subsampling rates, there are {considerable} {gains} in lateral resolution by the proposed DeepBF compared to its DAS, MVBF counterparts. {Again, similar qualitative performance improvement was seen using DeepBF and DeepMVBF,
but the contrast enhancement was noticeable using the proposed methods with deconvoluted targets (DeepDeconv).}
  This phenomenon was consistently observed for all sub-sampling factors.

{We believe that this may be originated from the inherent synthetic aperture that are originated from our training
that uses multiple scanlines, Rx, and depths as shown in \eqref{eq:train_new}. Recall that the input data $\sb_n$ is composed
of $\zb_{n-1},\zb_n,\zb_{n+1}$ which are composed of multiple scan line data as defined in \eqref{eq:zb}.
On the other hand, the resolution improvement along the axial directions was not significant,
 which may  be due to depth-independent training scheme in \eqref{eq:train_new}, which may lose the depth-dependent adaptivity. 
 The original training scheme in \eqref{eq:train_org} may be a solution for this, but requires huge memory.
 The way to overcome this technical limitation  is important, and will be investigated in other publications.}
 
{
In spite of the memory reduction using  \eqref{eq:train_new},
it still requires larger memory compared to the standard DAS, since it should  store multiple scan line data.  
In our future work we will explore the possible solutions to design an end-to-end beamformer method using a single-scanline data, which can learn the time delay but still provides better performance that the classical beamformers. 
In addition,  although the performance of the proposed deep learning approach is better than
the classical beamformer,  
after $2.7\times$ sub-sampling factor, the proposed method still degrades the image quality. Additional strategy to improve the performance of the proposed method is still required, which is the another important research direction.
Finally, the average reconstruction time of $4.8$ (milliseconds) per depth is still slow for real time implementation.
This is believed to be an engineering issue where
optimized software implementation other than Matlab in addition to use of multiple GPUs may address  the problem.
}

\begin{table*}[!hbt]
	\centering
	\caption{Performance statistics on \textit{in vivo} data for uniform sampling pattern}
	\label{tbl:results_sSTATS_invivo}
	\resizebox{0.9\textwidth}{!}{
		\begin{tabular}{c|ccc|ccc|ccc|ccc|ccc}
			\hline
			{sub-sampling} & \multicolumn{3}{c|}{{CR (dB)}} & \multicolumn{3}{c|}{{CNR}} & \multicolumn{3}{c|}{{GCNR}} & \multicolumn{3}{c|}{{PSNR (dB)}} & \multicolumn{3}{c}{{SSIM}}  \\
			{factor} & \textit{DAS} & \textit{MVBF} & \textit{DeepBF} & \textit{DAS} & \textit{MVBF} & \textit{DeepBF} & \textit{DAS} & \textit{MVBF} & \textit{DeepBF} & \textit{DAS} & \textit{MVBF} & \textit{DeepBF} & \textit{DAS} & \textit{MVBF} & \textit{DeepBF}  \\ \hline\hline
			1 & 12.37 & 12.39 & 13.25 & 1.38 & 1.38 & 1.45 & 0.64 & 0.64 & 0.66 & $\infty$ & $\infty$ & $\infty$ & 1 & 1 & 1 \\
			2 & 10.69 & 10.74 & 12.50 & 1.21 & 1.20 & 1.37 & 0.60 & 0.59 & 0.64 & 22.69 & 22.74 & 24.91 & 0.85 & 0.85 & 0.90 \\
			2.7 & 10.24 & 10.34 & 11.97 & 1.15 & 1.16 & 1.31 & 0.58 & 0.58 & 0.63 & 21.36 & 21.44 & 23.18 & 0.80 & 0.81 & 0.86 \\
			4& 9.67 & 9.80 & 11.00 & 1.10 & 1.10 & 1.22 & 0.56 & 0.56 & 0.60 & 20.08 & 20.19 & 21.38 & 0.75 & 0.76 & 0.80 \\
			8& 9.25 & 9.40 & 9.97 & 1.04 & 1.05 & 1.11 & 0.54 & 0.54 & 0.56 & 18.63 & 18.75 & 19.10 & 0.68 & 0.69 & 0.70 \\
			16& 9.08 & 9.25 & 9.58 & 1.02 & 1.03 & 1.08 & 0.53 & 0.53 & 0.55 & 17.84 & 17.95 & 17.83 & 0.63 & 0.63 & 0.64 \\ \hline
		\end{tabular}
	}
\end{table*}

\begin{table*}[!hbt]
	\centering
	\caption{Comparison of different design strategies with the proposed DeepBF on phantom dataset using random sub-sampling pattern}
	\label{tbl:results_selfcomp_phantom}
	\resizebox{0.9\textwidth}{!}{
		\begin{tabular}{c|cccc|cccc|cccc|cccc}
			\hline
			{Training/Design } & \multicolumn{4}{c|}{{CNR}} & \multicolumn{4}{c|}{{GCNR}} & \multicolumn{4}{c|}{{PSNR (dB)}} & \multicolumn{4}{c}{{SSIM}}  \\
			{methods} & 1$\times$ & 2$\times$ & 4$\times$ & 8$\times$ &
			1$\times$ & 2$\times$ & 4$\times$ & 8$\times$ &
			1$\times$ & 2$\times$ & 4$\times$ & 8$\times$ &
			1$\times$ & 2$\times$ & 4$\times$ & 8$\times$\\ \hline\hline
			\textit{RF-sum only} & 2.256 & 2.320 & 2.302 & 2.171 & 0.842 & 0.863 & 0.873 & 0.857 & $\infty$ & 25.37 & 21.12 & 18.62 & 1 & 0.924 & 0.832 & 0.737 \\
			\textit{fixed rate training} & 2.325 & 2.505 & 2.341 & 2.121 & 0.850   & 0.890 & 0.883 & 0.853 & $\infty$  & 22.30 & 21.30 & 19.88 & 1  & 0.880 & 0.840 & 0.777 \\
			\textit{single-depth design} & 2.752 & 2.659 & 2.505 & 2.253 & 0.913 & 0.907 & 0.895 & 0.873 & $\infty$ & 27.92 & 23.57 & 19.27 & 1 & 0.930 & 0.852 & 0.739 \\ \hline
			\textit{DeepBF(DAS)} & 2.655 & 2.650 & 2.527 & 2.248 & 0.897 & 0.900 & 0.898 & 0.874 & $\infty$ & 28.24 & 24.55 & 21.27 & 1 & 0.946 & 0.878 & 0.783 \\
			\textit{DeepBF(MVBF)} & 2.701 & 2.666 & 2.523 & 2.306 & 0.905 & 0.904 & 0.897 & 0.876 & $\infty$ & 29.20 & 25.17 & 22.45 & 1 & 0.952 & 0.890 & 0.812 \\
			\hline
		\end{tabular}
	}
\end{table*}


\section{Conclusion}
\label{sec:conclusion}

In this paper, we presented a novel deep learning-based adaptive and compressive beamformer to generate high-quality B-mode ultrasound image.
The proposed method is purely a data-driven method which exploits the spatio-temporal redundancies in the raw RF data, which help in generating improved quality B-mode images using various transducer numbers. 
{The proposed method can be trained using various targets, such as DAS, MVBF, and deconvoluted beamforming results from full RF data
to satisfy the desired IQ for each application.}
The proposed method improved the contrast of B-modes images by preserving the dynamic range and structural details of the RF signal in both the {phantom} and \textit{in vivo} scans.
Therefore, this method can be an important platform for ultrasound imaging.

\appendix
\setcounter{equation}{0}
\renewcommand{\theequation}{A.\arabic{equation}}

{Although this part is basically a summary   of  \cite{ye2019understanding}, we have included it
to make the paper self-contained.}

{Consider  encoder-decoder networks in Fig.~\ref{fig:CNN_block_diagram}.
The network has symmetric {configuration} so that
both encoder and decoder have the same number of layers, say $\kappa$;
the input and output dimensions for the encoder layer $\Ec^l$ and the decoder layer $\Dc^l$ are symmetric:
\begin{eqnarray*}
\Ec^l:\Rd^{d_{l-1}} \mapsto \Rd^{d_l},  \quad 
\Dc^l:\Rd^{d_{l}} \mapsto \Rd^{d_{l-1}}, \quad l\in[\kappa]
\end{eqnarray*}
where $[n]$ denoting the set $\{1,\cdots, n\}$.
 At the $l$-th layer, $m_l$ and $q_l$ denote the dimension of the signal, and  the number of filter channel, respectively. The length of filter
is assumed to be $r$.}

We now define the $l$-th layer input signal for the encoder layer from $q_{l-1}$ number of input channels
$$\zb^{l-1}:=\begin{bmatrix} \zb_1^{l-1\top} & \cdots & \zb^{l-1\top}_{q_{l-1}} \end{bmatrix}^\top \in   \Rd^{d_{l-1}}, \quad  $$ 
where  $^\top$ denotes the transpose,
and $\zb_j^{l-1} \in \Rd^{m_{l-1}}$ refers to the $j$-th channel input with the dimension $m_{l-1}$.
The $l$-th layer output signal $\zb^l$ is similarly defined.
Then, 
we have the following representation of the convolution and pooling operation at the $l$-th encoder layer \cite{ye2019understanding}:
\begin{eqnarray}\label{eq:Enc}
 \z^l
 = \sigma\left(\Eb^{l\top} \z^{l-1}\right)
\end{eqnarray}
where $\sigma(\cdot)$ is defined as an element-by-element ReLU operation $\sigma(x)=\max\{x,0\}$, and 
   \begin{eqnarray}\label{eq:El}
\E^l= \begin{bmatrix} 
\Phib^l\circledast \psib^l_{1,1} & \cdots &  \Phib^l\circledast \psib^l_{q_l,1}  \\
  \vdots & \ddots & \vdots \\
\Phib^l\circledast \psib^l_{1,q_{l-1}} & \cdots &\Phib^l\circledast \psib^l_{q_{l},q_{l-1}}
 \end{bmatrix}
 \end{eqnarray}
 where $\Phib^l$ denotes the $m_l\times m_l$  matrix that represents the pooling operation at the $l$-th layer, and $\psib_{i,j}^l\in\Rd^C$ represents
the $l$-th layer encoder filter to generate the $i$-th channel output from the contribution of the $j$-th channel input,
and
 $\circledast$ represents a multi-channel convolution \cite{ye2019understanding}.

Similarly, the $l$-th decoder layer can be represented by  \cite{ye2019understanding}:
\begin{eqnarray}\label{eq:Dec}
\tilde \z^{l-1}=\sigma\left(\Db^l \tilde\z^{l} \right)
\end{eqnarray}
where 
   \begin{eqnarray}
 \Db^l= \begin{bmatrix} 
\tilde\Phib^l\circledast \tilde\psib^l_{1,1} & \cdots &  \tilde\Phib^l \circledast \tilde\psib^l_{1,q_l}  \\
  \vdots & \ddots & \vdots \\
\tilde\Phib^l\circledast \tilde\psib^l_{q_{l-1},1} & \cdots &  \tilde\Phib^l\circledast \tilde\psib^l_{q_{l-1},q_{l}}
 \end{bmatrix}
 \end{eqnarray}
 where 
$\tilde\Phib^l$ denotes the $m_l\times m_l$  matrix that represents the unpooling operation at the $l$-th layer, and
  $\tilde\psib_{i,j}^l\in \Rd^r$ represents
the $l$-th layer decoder filter to generate the $i$-th channel output from the contribution of the $j$-th channel input.

Next, consider  the skipped branch signal $\chib^l$ by concatenating the output for each skipped branch as shown in Fig.~\ref{fig:CNN_block_diagram}.
Then,  the $l$-th encoder layer with the skipped connection 
can be represented
by  \cite{ye2019understanding}:
\begin{align}
\begin{bmatrix}\z^l\\ \chib^l \end{bmatrix} 
 &= \begin{bmatrix} \sigma\left(\Eb^{l\top} \z^{l-1}\right)\\  \z^{l-1}\end{bmatrix} \label{eq:skip_enc}\\
 \tilde \z^{l-1} &=\sigma\left(\Db^l \tilde\z^{l} +\Db^l \chib^l\right) \ . \label{eq:skip_dec}
\end{align}

With these definition,
by following the derivation {in} \cite{ye2019understanding}, it is straightforward to show that
the neural network output $\vb$ is given by
\begin{eqnarray}\label{eq:basis}
\vb  &=& \tilde\Bb(\zb)\Bb(\zb)\zb = \sum_{i} \left\langle {\blmath b}_i(\z), \z \right\rangle \tilde  {\blmath b}_i(\z)
\end{eqnarray}
where  $\Thetab$ refers to all the convolution filter parameters,
and $ {\blmath b}_i(\z)$ and $\tilde  {\blmath b}_i(\z)$ denote the $i$-th column of the matrices \eqref{eq:Bz} and \eqref{eq:tBz}, respectively, which are defined as

\begin{eqnarray}
\B(\z) &=& \left[ \Eb^1\Lambdab^1(\z)\Eb^2 \cdots  \Lambdab^{\kappa-1}(\z) \Eb^{\kappa}, \cdots \right.  \label{eq:Bz} \\
&& \left. \cdots, \Eb^1\Lambdab^1(\z)\cdots\Eb^6,~  \Eb^1\Lambdab^1(\z)\cdots\Eb^3 \right] \notag
\end{eqnarray}
\begin{eqnarray}
\tilde\B(\z) &=& \left[ \Db^1\tilde\Lambdab^1(\z)\Db^2 \cdots  \tilde\Lambdab^{\kappa-1}(\z) \Db^{\kappa}, \cdots \right. \label{eq:tBz}\\
&& \left. \cdots, \Db^1\tilde\Lambdab^1(\z)\cdots\Db^6,~  \Db^1\tilde\Lambdab^1(\z)\cdots\Db^3 \right] \notag
\end{eqnarray}
where
$\Lambdab^l(\z)$ and $\tilde\Lambdab^l(\z)$ denote the diagonal matrix with 0 and 1 values that are determined by the ReLU output
in the previous convolution steps. Note that 
 there are skipped connections at every third convolution operations in Fig.~\ref{fig:CNN_block_diagram}, so that
 the last blocks in \eqref{eq:Bz} and \eqref{eq:tBz} are indexed accordingly.

\bibliographystyle{IEEEtran}
\bibliography{Final_Version}

\end{document}